\PassOptionsToPackage{unicode}{hyperref}
\PassOptionsToPackage{hyphens}{url}
\PassOptionsToPackage{dvipsnames,svgnames,x11names}{xcolor}

\documentclass[
  12pt]{article}

\usepackage{wrapfig}
\usepackage{siunitx}
\usepackage{enumitem}

\usepackage{amsmath,amssymb}
\usepackage{iftex}
\ifPDFTeX
  \usepackage[T1]{fontenc}
  \usepackage[utf8]{inputenc}
  \usepackage{textcomp} 
\else 
  \usepackage{unicode-math}
  \defaultfontfeatures{Scale=MatchLowercase}
  \defaultfontfeatures[\rmfamily]{Ligatures=TeX,Scale=1}
\fi
\usepackage{lmodern}
\ifPDFTeX\else  
\fi
\IfFileExists{upquote.sty}{\usepackage{upquote}}{}
\IfFileExists{microtype.sty}{
  \usepackage[]{microtype}
  \UseMicrotypeSet[protrusion]{basicmath} 
}{}
\makeatletter
\@ifundefined{KOMAClassName}{
  \IfFileExists{parskip.sty}{%
    \usepackage{parskip}
  }{
    \setlength{\parindent}{0pt}
    \setlength{\parskip}{6pt plus 2pt minus 1pt}}
}{
  \KOMAoptions{parskip=half}}
\makeatother
\usepackage{xcolor}
\makeatletter
\ifx\paragraph\undefined\else
  \let\oldparagraph\paragraph
  \renewcommand{\paragraph}{
    \@ifstar
      \xxxParagraphStar
      \xxxParagraphNoStar
  }
  \newcommand{\xxxParagraphStar}[1]{\oldparagraph*{#1}\mbox{}}
  \newcommand{\xxxParagraphNoStar}[1]{\oldparagraph{#1}\mbox{}}
\fi
\ifx\subparagraph\undefined\else
  \let\oldsubparagraph\subparagraph
  \renewcommand{\subparagraph}{
    \@ifstar
      \xxxSubParagraphStar
      \xxxSubParagraphNoStar
  }
  \newcommand{\xxxSubParagraphStar}[1]{\oldsubparagraph*{#1}\mbox{}}
  \newcommand{\xxxSubParagraphNoStar}[1]{\oldsubparagraph{#1}\mbox{}}
\fi
\makeatother

\usepackage{longtable,booktabs,array}
\usepackage{calc} 
\usepackage{etoolbox}
\makeatletter
\patchcmd\longtable{\par}{\if@noskipsec\mbox{}\fi\par}{}{}
\makeatother
\IfFileExists{footnotehyper.sty}{\usepackage{footnotehyper}}{\usepackage{footnote}}
\makesavenoteenv{longtable}
\usepackage{graphicx}
\makeatletter
\def\maxwidth{\ifdim\Gin@nat@width>\linewidth\linewidth\else\Gin@nat@width\fi}
\def\maxheight{\ifdim\Gin@nat@height>\textheight\textheight\else\Gin@nat@height\fi}
\makeatother
\setkeys{Gin}{width=\maxwidth,height=\maxheight,keepaspectratio}
\makeatletter
\def\fps@figure{htbp}
\makeatother

\makeatletter
\@ifpackageloaded{caption}{}{\usepackage{caption}}
\AtBeginDocument{%
\ifdefined\contentsname
  \renewcommand*\contentsname{Table of contents}
\else
  \newcommand\contentsname{Table of contents}
\fi
\ifdefined\listfigurename
  \renewcommand*\listfigurename{List of Figures}
\else
  \newcommand\listfigurename{List of Figures}
\fi
\ifdefined\listtablename
  \renewcommand*\listtablename{List of Tables}
\else
  \newcommand\listtablename{List of Tables}
\fi
\ifdefined\figurename
  \renewcommand*\figurename{Figure}
\else
  \newcommand\figurename{Figure}
\fi
\ifdefined\tablename
  \renewcommand*\tablename{Table}
\else
  \newcommand\tablename{Table}
\fi
}
\@ifpackageloaded{float}{}{\usepackage{float}}
\floatstyle{ruled}
\@ifundefined{c@chapter}{\newfloat{codelisting}{h}{lop}}{\newfloat{codelisting}{h}{lop}[chapter]}
\floatname{codelisting}{Listing}

\makeatother
\makeatletter
\@ifpackageloaded{caption}{}{\usepackage{caption}}
\@ifpackageloaded{subcaption}{}{\usepackage{subcaption}}
\makeatother

\ifLuaTeX
  \usepackage{selnolig}  
\fi
\usepackage{natbib}
\usepackage{bookmark}

\IfFileExists{xurl.sty}{\usepackage{xurl}}{} 
\hypersetup{
  pdftitle={Scalable Spatial Stream Network (S3N) Models},
  pdfauthor={Jessica P. Kunke; Julian D. Olden; Tyler H. McCormick},
  pdfkeywords={SSNs, spatial processes, ecology, networks, computational efficiency, nearest-neighbor Gaussian processes},
  colorlinks=true,
  linkcolor={blue},
  filecolor={Maroon},
  citecolor={Blue},
  urlcolor={Blue},
  pdfcreator={LaTeX via pandoc}}

\newcommand{\anon}{1}

\begin{document}


\def\spacingset#1{\renewcommand{\baselinestretch}%
{#1}\small\normalsize} \spacingset{1}


\if1\anon
{
  \title{\bf Scalable Spatial Stream Network (S3N) Models}
  \author{Jessica P. Kunke\footnote{Correspondence: Jessica P. Kunke, jess.kunke@montana.edu}\\
    Department of Mathematical Sciences, Montana State University\\
    Julian D. Olden \\
    School of Aquatic and Fishery Sciences, University of Washington\\
    and \\
    Tyler H. McCormick \\
    Department of Statistics, University of Washington}
  \maketitle
} \fi

\if0\anon
{
  \bigskip
  \bigskip
  \bigskip
  \begin{center}
    {\LARGE\bf Scalable Spatial Stream Network (S3N) Models}
\end{center}
  \medskip
} \fi

\bigskip
\begin{abstract}
Understanding how habitats shape species distributions and abundances across river networks remains a longstanding and fundamental challenge in ecology, with direct implications for effective biodiversity management and conservation. We introduce a scalable spatial stream network (S3N) model that enables estimation, inference, and prediction with greater computational efficiency than existing spatial stream network software. S3Ns extend nearest-neighbor Gaussian processes (NNGPs) to include ecologically salient stream network dependence structure. S3Ns also streamline preprocessing, whose cost depends on both the number of reaches and the number of observations. We demonstrate that S3Ns recover fixed effect parameters as accurately as existing spatial stream network models while running up to 2--3 orders of magnitude faster on large networks. We then apply S3Ns to estimate the population sizes and geographic distributions of 285 fish species in the entire Ohio River Basin (approximately 170,000 reaches and 9,000 observation points) on a laptop, with pairwise distance computations run on a computing cluster. These results indicate the promise of S3Ns for mapping freshwater variables and quantifying the influence of environmental drivers across extensive, complex river networks with many observation points.
\end{abstract}

\noindent%
{\it Keywords:} SSNs, spatial processes, ecology, networks, computational efficiency, nearest-neighbor Gaussian processes
\vfill

\newpage
\spacingset{1.75}

\section{Introduction}\label{sec:intro}

Freshwater ecosystems are among the most diverse on Earth, providing essential benefits such as fisheries, clean drinking water, material resources, and cultural value to billions of people \citep{Lynch2023}. Despite covering a small fraction of the planet’s surface, fresh waters also support roughly 12\% of all described animal species, including nearly one-third of all vertebrates \citep{Darwall2018}. Yet, freshwater biodiversity is facing severe and accelerating declines\textemdash vertebrate populations in freshwater ecosystems are decreasing at nearly four times the rate of those in marine or terrestrial environments and have greater future projected extinction rates \citep{Tickner2020_GlobalFWBiodiversityLoss, Urban2024_climatechange}. Freshwater species face intense pressure from a wide range of persistent and emerging threats, including climate change, land-use modification, pollution, invasive species, infectious diseases, algal blooms, flow alteration, over-exploitation, salinization, and contaminants such as pesticides, microplastics, and pharmaceuticals \citep{Reid2019_emerging}. To facilitate management interventions that can curtail or reverse the decline in freshwater biodiversity, research must continue to address key knowledge gaps that impede progress \citep{Tickner2020_GlobalFWBiodiversityLoss, Harper2021}.

A central challenge in freshwater ecology is to predict how present and future habitat conditions influence species distributions and population abundances, and to apply these insights to inform effective fisheries management \citep{Paukert2021}. Modeling the distribution of freshwater fishes involves various approaches that seek to capture the complex interplay between environmental conditions, biotic interactions and the spatial structure of rivers that is unique to freshwater ecosystems. Species distribution models are widely used to predict suitable habitats by correlating species occurrences with environmental variables \citep[e.g.,][]{Bond2011, Markovic2012, Radinger2017}, in addition to accounting for species interactions \citep[e.g.,][]{Wagner2020}. More spatially explicit methods, such as network-based and graph-theoretical models, consider the dendritic structure of river systems and the effects of connectivity on fish distributions \citep{Eros2012characterizing}. Process-based and individual-based models simulate ecological dynamics like dispersal, reproduction, and survival, offering a mechanistic view of species responses to changing conditions \citep{Rogosch2019, Tonkin2021}. The choice of model depends on the research question, spatial grain and extent, data availability, computational limitations, and the desire to incorporate river network structure and the contributing upstream watershed.

Accounting for the branching structure of rivers offers a more robust understanding of the spatial dynamics shaping riverine biodiversity \citep{Fausch2002, CampbellGrant2007, Altermatt2013}, supporting better management and conservation practices. Spatial dependence among observations along a river network is a function not only of Euclidean (bird's eye) distance, but also of the directed or undirected distance along the stream network \citep{Dent1999_StreamWaterNutrient, Wyatt2003_Mapping, Ganio2005_GeostatApproach, Peterson2006_StreamWaterChemistry, Webster2008_Bayesian}. For instance, waterborne chemicals and larval invertebrates may passively flow downstream, while fish can actively swim downstream or upstream at rates that vary depending on the species \citep[e.g.,][]{Bilton2001_dispersal, Schofield2018, Leibowitz2018, Comte2018_fish}. \cite{VerHoef2006_SpatialModelsThatUseFlow} demonstrated that substituting stream distance for Euclidean distance in existing covariance functions does not guarantee a valid (positive definite) covariance function. \cite{VerHoef2010_MovingAverageApproach} extended Gaussian process (GP) models to the stream network context by using moving-average constructions to construct covariance functions based on stream distance. These spatial stream network (SSN) models were first implemented in the \texttt{STARS} ArcGIS toolkit and the original \texttt{SSN} R package \citep{Peterson2014_STARS, VerHoef2014_SSN}, which have since been superseded by the \texttt{SSNbler} and \texttt{SSN2} R packages for preprocessing and model fitting, respectively \citep{Peterson2024_SSNbler, Dumelle2024_SSN2}. SSNs revolutionized how scientists model ecological processes and organisms in river networks.

Unfortunately, exact SSN models quickly become computationally prohibitive. Given $n$ observation locations, fitting exact spatial process models requires $O(n^3)$ floating point operations (flops) and $O(n^2)$ storage to compute the inverse and determinant of the full $n\times n$ covariance matrix \citep{RasmussenWilliams2006_GPs, Datta2016_NNGP_Interpretation}. Fitting SSNs additionally requires computing pairwise stream distances and other stream network information in order to compute covariances; we and previous authors refer to this as \textbf{preprocessing}. While calculating the Euclidean distance between two points requires only their Cartesian coordinates, computing the stream distance between two points requires knowledge of stream network structure between the two points\textemdash whether water flows directly from one point to another, or from both points to meet at some common junction further downstream. Computing these path lengths requires knowledge of how different river segments, called \textbf{reaches}, are connected. While the computational expense of model fitting is a function only of the number of observation locations, the expense of preprocessing increases with both the number of reaches and observation locations. 

In this paper, we present the scalable spatial stream network (S3N) model to address the computational expense of both the preprocessing and model fitting. S3Ns extend an approximation known as nearest-neighbor Gaussian processes (NNGPs) to a stream context, reducing the computational complexity of model fitting from $O(n^3)$ to $O(n)$. We use the stream network structure to choose the nearest neighbors. To reduce preprocessing expense, we leverage the adjacency matrices of the stream network graph and compute stream network quantities by starting at the stream outlet and working up the network one network level at a time.  For large data sets, \texttt{SSN2} offers a big-data approximation based on spatial partitioning \citep{VerHoef2023_IndexingSPIN}: observations are divided into groups of about 200, by default using k-means clustering on their spatial coordinates, and dependence between groups is ignored during estimation. S3Ns take a different approach. Rather than partitioning the data, an S3N conditions every observation on a small set of informative neighbors chosen by stream distance and flow connectivity, yielding a single valid spatial process over the whole network. We compare S3N against \texttt{SSN2} with this big-data approximation, the fastest existing option for fitting SSNs to large data sets, and demonstrate that model fitting can be up to 54 times as fast with S3Ns, and preprocessing up to two to three orders of magnitude faster on large networks.

We begin with a detailed description of the S3N model (Section \ref{sec:S3N-models}), then present simulation studies that confirm S3Ns provide essential computational gains over existing models without compromising accuracy, permitting the analysis of larger networks with more observations (Section \ref{sec:benchmarking-validation}). Next, as a demonstration on real data, we apply the S3N model to estimate the geographical distributions, reach-scale densities, and total population sizes for observations of 285 fish species at 8,924 locations across the Ohio River Basin, one of the most species-rich river systems in North America with diverse ecological habitats (Section \ref{sec:case-study}). Lastly, we summarize the findings and limitations of this study and highlight directions for future work (Section \ref{sec:discussion}).

\section{Scalable spatial stream network models (S3Ns)}\label{sec:S3N-models}

Let $\boldsymbol{Y} = (Y(\boldsymbol{s_1}), ..., Y(\boldsymbol{s_n}))$ be the observed fish densities (the number of fish per 100-m length of river) at $n$ point locations $\boldsymbol{s_1}$, ..., $\boldsymbol{s_n}$ along a stream network. Given these fish densities, we wish to predict or interpolate the response variable across the entire regional stream network and estimate region-wide summary statistics. The fundamental basis for the S3N model is a Gaussian process (GP), under which $\boldsymbol{Y}$ has a multivariate normal distribution whose mean is a linear function of $p$ covariates (represented in matrix form as $\boldsymbol{X}$ with coefficients $\boldsymbol\beta$) and whose covariance is given by a covariance function $\Sigma$ \citep{CressieWikle2011_SpatiotemporalData}: \begin{equation} 
        \boldsymbol{Y} = \boldsymbol{X}\boldsymbol\beta + \boldsymbol{w}, \quad \boldsymbol{w} \sim N\left(0, \boldsymbol{\Sigma}\right), \quad \Sigma\big(\boldsymbol{s_i}, \boldsymbol{s_j}\big) = C(\boldsymbol{s_i}, \boldsymbol{s_j}) + \tau^2 \delta_{ij}.
\label{eq-GP-model}
\end{equation}

Here, $\tau^2 \delta_{ij}$ captures any independent error, whether from measurement error or from variation at scales smaller than the distances among observations, and $C(\boldsymbol{s_i}, \boldsymbol{s_j})$ represents the spatial covariance. Outside the stream context, $C(\boldsymbol{s_i}, \boldsymbol{s_j})$ is often parameterized as a function of the Euclidean distance between the two points: $C(\boldsymbol{s_i}, \boldsymbol{s_j}) = C_e(\boldsymbol{s_i}, \boldsymbol{s_j} \mid \boldsymbol\theta_e)$, where $\boldsymbol\theta_e$ represents the parameters of this function.

Estimating the parameters of a GP requires storing the covariance matrix and computing its inverse and determinant. The full covariance matrix for estimation with $n$ observation point locations is $n\times n$. Computing the full likelihood for estimation therefore requires $O(n^2)$ storage, and computing the inverse and determinant of the covariance matrix using the Cholesky decomposition requires $O(n^3)$ floating point operations (flops). Many different methods have been developed to provide fast approximations of spatial process models, including Vecchia's nearest-neighbor approximation. The main alternative class of approaches, low rank models, are still too computationally expensive for large $n$ and perform poorly in the presence of high spatial correlation, both of which are critical for regression on large spatial datasets \citep{Stein2014_lowrank_limitations, Datta2016_NNGP}. We focus on NNGPs since these extend the Vecchia approximation into valid generative models in their own right, enabling prediction at arbitrary locations aside from the observations; further details are provided below. Additionally, nearest-neighbor approximations are expected to be most accurate when spatial dependence is concentrated at short distances, since conditioning on nearby observations then captures most of the information in the full process \citep{Stein2002_screening}. This is plausible for many freshwater fish, whose movement and dispersal are limited relative to the extent of large river networks.

\subsection{NNGPs}\label{sec:NNGPs}

NNGPs are based on the sparse nearest-neighbor approximation proposed by \cite{Vecchia1988_sparsity}. The full likelihood\textemdash the joint probability of the values of the GP at all observation points\textemdash can be expressed as a product of conditional densities using the general product rule: \vspace{-1em}\begin{align}
    p(\boldsymbol{Y}) = &p\big(Y(\boldsymbol{s}_1)\big) \cdot p\big(Y(\boldsymbol{s}_2)\mid Y(\boldsymbol{s}_1)\big) \cdots p\big(Y(\boldsymbol{s}_n)\mid Y(\boldsymbol{s}_{n-1}), \ldots , Y(\boldsymbol{s}_1)\big) \nonumber\\
    = & p\big(Y(\boldsymbol{s}_1)\big) \prod_{i=2}^n p\big(Y(\boldsymbol{s}_i)\mid \boldsymbol{Y}(H(\boldsymbol{s}_i))\big),
    \label{eq-product-rule}
\end{align}

where $H(\boldsymbol{s}_i) := \{\boldsymbol{s}_{i-1}, \ldots , \boldsymbol{s}_1\}$ is the conditioning set for $\boldsymbol{s}_i$ and $\boldsymbol{Y}(H(\boldsymbol{s}_i))$ is the vector of responses (fish densities, in our case) at the locations in $H(\boldsymbol{s}_i)$. Equation \ref{eq-product-rule} holds for any ordering of the locations $s_i$. \cite{Vecchia1988_sparsity} proposed approximating the likelihood by conditioning each point on at most some fixed number $m$ of its nearest neighbors, replacing $H(\boldsymbol{s}_i)$ with a subset $M(\boldsymbol{s}_i) \subseteq H(\boldsymbol{s}_i)$ of $m$ or fewer points that are nearest to $\boldsymbol{s}_i$ by Euclidean distance. Specifically, for $i \le m$, $M(\boldsymbol{s}_i) = H(\boldsymbol{s}_i)$ and $|M(\boldsymbol{s}_i)| = i \le m$. For $i>m$, $M(\boldsymbol{s}_i)$ is the set of $m$ nearest neighbors by Euclidean distance to $\boldsymbol{s}_i$, and $|M(\boldsymbol{s}_i)| = m$. The resulting likelihood approximation is \begin{equation}
    p(\boldsymbol{Y}) \approx p\big(Y(\boldsymbol{s}_1)\big) \prod_{i=2}^n p\big(Y(\boldsymbol{s}_i)\mid \boldsymbol{Y}(M(\boldsymbol{s}_i))\big).
    \label{eq-Vecchia-approx}
\end{equation} 
This approximate likelihood only requires storing and computing $n$ matrices of size at most $m\times m$, reducing storage from $O(n^2)$ to $O(nm^2)$ and computational cost in flops from $O(n^3)$ to $O(nm^3)$.

A limitation of likelihood approximation methods such as that of \cite{Vecchia1988_sparsity} in Equation \ref{eq-Vecchia-approx} above is that they define an approximation at the observed locations; they do not necessarily allow us to predict values at other locations unless they are shown to correspond to some underlying process. \cite{Datta2016_NNGP} addressed this limitation by proving that Vecchia's approximation can be extended from a likelihood approximation to a valid spatial process, a nearest-neighbor Gaussian process (NNGP), which approximates the original GP. The NNGP model is given by \begin{equation}
        \boldsymbol{Y} = \boldsymbol{X}\boldsymbol\beta + \boldsymbol{w}, \quad \boldsymbol{w} \sim N\left(0, \boldsymbol{\tilde{\Sigma}}\right), \quad w(\boldsymbol{s}_i) \mid \boldsymbol{w}(M(\boldsymbol{s}_i)) \sim N(\boldsymbol{a_i}^T\boldsymbol{w}(M(\boldsymbol{s}_i)), d_i),
        \nonumber
    \end{equation}
    \begin{equation}
        \boldsymbol{a_i} = \boldsymbol{\Sigma}\big(M(\boldsymbol{s_i}), M(\boldsymbol{s_i})\big)^{-1} \boldsymbol{\Sigma}\big(M(\boldsymbol{s_i}), \boldsymbol{s_i}\big), \quad d_i = \Sigma\big(\boldsymbol{s_i}, \boldsymbol{s_i}\big) - \boldsymbol{\Sigma}\big(\boldsymbol{s_i}, M(\boldsymbol{s_i})\big)\boldsymbol{a_i},
        \label{eq-NNGP-model}
    \end{equation}
    \begin{equation}
        \Sigma\big(\boldsymbol{s_i}, \boldsymbol{s_j}\big) =  C_e(\boldsymbol{s_i}, \boldsymbol{s_j} \mid \boldsymbol\theta_e) + \tau^2 \delta_{ij}.
        \nonumber
\end{equation}

We write $\boldsymbol{w} \sim \mathrm{NNGP}(\Sigma, M)$ for the process defined by Equation \ref{eq-NNGP-model}, which is fully determined by the parent covariance function $\Sigma$ and the neighbor sets $M = \{M(\boldsymbol{s}_1), \ldots, M(\boldsymbol{s}_n)\}$. The covariance matrix of this approximating NNGP is $\boldsymbol{\tilde{\Sigma}} := (\boldsymbol{I}-\boldsymbol{A})^{-1}\boldsymbol{D}(\boldsymbol{I}-\boldsymbol{A})^{-T}$, where $\boldsymbol{A} = (a_{ij})$ is a strictly lower triangular matrix, $\boldsymbol{D} = \text{diag}(d_1, ..., d_n)$, and $\boldsymbol{I}$ is the identity matrix. The Cholesky factor of the precision matrix $\boldsymbol{\tilde{\Sigma}}^{-1}$, $\boldsymbol{\tilde{L}} = \boldsymbol{D}^{-1/2}(\boldsymbol{I}-\boldsymbol{A})$, requires only $O(nm^3)$ flops and $O(nm^2)$ storage, compared to $O(n^3)$ flops and $O(n^2)$ storage for the full original process.

\subsection{The S3N model}\label{sec:S3Nmodel}

Vecchia-type approximations have been developed for non-Euclidean distances \citep{KangKatzfuss2023_correlation}, but to our knowledge, NNGPs have not yet been implemented for data with an underlying network structure. The network structure can affect both which neighbors are nearest or most informative and how the covariance function is defined. In the context of streams, define the upstream distance $U(\boldsymbol{s}_i)$ of a point $\boldsymbol{s}_i$ on the stream network to be the length of the flow path along the network from $\boldsymbol{s}_i$ to its stream outlet. The stream distance $h(\boldsymbol{s}_i, \boldsymbol{s}_j)$ between two points $\boldsymbol{s}_i$ and $\boldsymbol{s}_j$ can then be computed as \[h(\boldsymbol{s}_i, \boldsymbol{s}_j) = 
\begin{cases} 
    \lvert U(\boldsymbol{s}_i) - U(\boldsymbol{s}_j) \rvert & \text{if } \boldsymbol{s}_i, \boldsymbol{s}_j \text{ are flow-connected,}\\
    U(\boldsymbol{s}_i) + U(\boldsymbol{s}_j) - 2U(\boldsymbol{c}_{ij}) & \text{else,}
\end{cases}\] where $\boldsymbol{c}_{ij}$ is the nearest common junction downstream of points $\boldsymbol{s}_i$ and $\boldsymbol{s}_j$. 

Whereas \cite{Vecchia1988_sparsity} and \cite{Saha2018_BRISC} choose the neighbor sets $M(s_i)$ to be the $m$ nearest neighbors of $s_i$ among $s_1, ..., s_{i-1}$ with respect to Euclidean distance, we instead choose the $m$ nearest neighbors using stream distance and flow connectivity (see Section \ref{sec:preprocessing}), and we include covariance components that are a function of stream distance: \begin{equation}
        \Sigma_s\big(\boldsymbol{s_i}, \boldsymbol{s_j}\big) = \pi_{ij} C_u(\boldsymbol{s_i}, \boldsymbol{s_j} \mid \boldsymbol\theta_u) + C_d(\boldsymbol{s_i}, \boldsymbol{s_j} \mid \boldsymbol\theta_d) + C_e(\boldsymbol{s_i}, \boldsymbol{s_j} \mid \boldsymbol\theta_e) + \tau^2 \delta_{ij}.
        \label{eq-S3N-model}
\end{equation}

Here, $C_e$ represents a standard covariance function based on Euclidean distance while $C_u$ and $C_d$ represent the two classes of stream covariance functions, tail-up and tail-down, developed by \cite{VerHoef2010_MovingAverageApproach} using the approach from \cite{Yaglom1987_MAcovar} and \cite{VerHoefBarry1996_MAcovar} of convolving moving average functions with a white noise process. For tail-up covariance functions, the moving average function is non-zero only upstream of the point in question, while for tail-down covariance functions, the moving average function is non-zero only downstream of the point in question.

Spatial correlation between two points occurs if and only if their moving average functions overlap. As a result, spatial dependency with tail-up covariance occurs between two points only if the stream flows from one point to the other (flow-connected points), while both flow-connected and flow-unconnected pairs of points can be spatially correlated with tail-down covariance. Tail-up models are therefore useful for modeling waterborne chemicals and small organisms that move passively with the stream flow, as well as fish that can swim more easily downstream. The spatial weights $\pi_{ij}$ account for the proportional influence of confluent stream segments; see \cite{VerHoef2010_MovingAverageApproach} and \cite{VerHoef2019_SpatialStreamModels} for further details. Note that for points that are not flow-connected, tail-down covariances sometimes also require the two distances from each point to their nearest common junction separately. In this case, define $a(\boldsymbol{s}_i, \boldsymbol{s}_j) = \min\big(U(\boldsymbol{s}_i), U(\boldsymbol{s}_j)\big) - U(\boldsymbol{c}_{ij})$ to be the shorter of these two distances and $b(\boldsymbol{s}_i, \boldsymbol{s}_j) = \max\big(U(\boldsymbol{s}_i), U(\boldsymbol{s}_j)\big) - U(\boldsymbol{c}_{ij})$ to be the longer.

Combining these two changes, the S3N model is
\begin{equation}
    \boldsymbol{Y} = \boldsymbol{X}\boldsymbol\beta + \boldsymbol{w}, \qquad
    \boldsymbol{w} \sim \mathrm{NNGP}\big(\Sigma_s, M_s\big),
    \label{eq-S3N-def}
\end{equation}
where $\Sigma_s$ is the stream covariance in Equation \ref{eq-S3N-model} and $M_s$ are neighbor sets chosen using stream distance and flow connectivity (Section \ref{sec:preprocessing}). Because the NNGP construction yields a valid Gaussian process for any valid parent covariance function and any choice of neighbor sets \citep{Datta2016_NNGP}, and the tail-up, tail-down, and Euclidean components are each valid covariance functions on a stream network \citep{VerHoef2010_MovingAverageApproach}, the S3N is itself a valid Gaussian process. Predictions at unobserved locations are therefore well defined. The stream-based neighbor sets also matter for the quality of the approximation: under tail-up covariance, a flow-unconnected neighbor has zero covariance with the location it conditions and so contributes nothing to its conditional distribution. Choosing neighbors by Euclidean distance, as standard NNGP software does, could therefore leave many locations with uninformative neighbor sets. Unlike partitioning approaches such as the spatial indexing approach of \citet{VerHoef2023_IndexingSPIN}, which ignore dependence between groups and form groups by Euclidean rather than stream proximity, the NNGP construction keeps every observation conditioned on nearby flow-connected neighbors.

\subsection{Preprocessing}\label{sec:preprocessing}

Computing the covariance in Equation \ref{eq-S3N-model} requires, for every observation and prediction location, its upstream distance, its position in the network topology, and the additive function value (AFV) used to compute the spatial weights $\pi_{ij}$.
S3N computes these quantities in four steps; we refer to the first three as \textbf{network configuration} and the fourth as \textbf{pairwise distance computation}, and to all four together as preprocessing.

\textbf{Step 1: Building the network graph.} Each reach is treated as a polyline digitized in the direction of flow, so its first and last vertices are its upstream and downstream nodes, respectively. We construct a directed graph whose vertices are these nodes and whose edges are the reaches, together with its line graph, in which each reach is a vertex and a directed edge joins reach $k$ to reach $l$ if $k$ flows directly into $l$. We store the adjacency matrix $\boldsymbol{J}$ of the line graph, where $J_{kl} = 1$ if reach $k$ flows into reach $l$ and $J_{kl} = 0$ otherwise, as a sparse matrix. Its row and column sums expose the network topology: a row sum of 0 identifies a sink (a stream outlet) and a row sum greater than 1 a downstream divergence, while a column sum of 0 identifies a source and a column sum greater than 2 a complex confluence. Divergences and complex confluences must be corrected by the user before the next step. The connected components of the graph identify separate networks, each with its own outlet. S3N does not currently snap nearly coincident endpoints or detect dangling nodes or intersections without nodes, but these features can be added. S3N's checks for divergences and complex confluences apply to any stream network, but networks should currently be corrected for these other error types before using S3N; the NSI flowlines used here have already been corrected for them, although they still contained a complex confluence that S3N detected (Section \ref{sec:case-study}).

\textbf{Step 2: Computing stream variables working upstream from the outlet.} Let $U_k$ denote the upstream distance of the downstream node of reach $k$, $L_k$ its length, and $W_k$ its total upstream drainage area. Starting from the outlet reaches, for which $U_k = 0$ and $\mathrm{AFV}_k = 1$, each iteration uses the nonzero entries of $\boldsymbol{J}$ to find every reach $k$ that flows into a reach $l$ processed in the previous iteration, and sets
\begin{equation}
    U_k = U_l + L_l, \qquad
    \omega_k = \frac{W_k}{\sum_{k' : J_{k'l} = 1} W_{k'}}, \qquad
    \mathrm{AFV}_k = \omega_k \, \mathrm{AFV}_l ,
    \label{eq-updist-recursion}
\end{equation}
where $\omega_k$ is the segment proportional influence of reach $k$. Each reach also receives a binary identifier formed by appending a 0 or 1 to the identifier of the reach it flows into, so that the identifier encodes the full flow path from the outlet \citep{Peterson2014_STARS, VerHoef2014_SSN}. All reaches at the same depth are processed together in vectorized operations, so each reach is visited exactly once and the number of iterations equals the largest number of reaches along any flow path, rather than the total number of reaches.

\textbf{Step 3: Adding observation and prediction locations.} Each location inherits the binary identifier and AFV of the reach on which it lies. Its upstream distance is the upstream distance of that reach's downstream node plus the distance along the reach from that node to the location.

\textbf{Step 4: Pairwise distance computation and neighbor selection.} Because binary identifiers encode flow paths, the nearest common junction $\boldsymbol{c}_{ij}$ of locations $\boldsymbol{s}_i$ and $\boldsymbol{s}_j$ corresponds to the longest common prefix of their identifiers, and the two locations are flow-connected exactly when that prefix equals one of the two identifiers. The stream distance $h(\boldsymbol{s}_i, \boldsymbol{s}_j)$ then follows from the expression above, and the tail-up spatial weight is $\pi_{ij} = \sqrt{\mathrm{AFV}_{\text{up}} / \mathrm{AFV}_{\text{down}}}$ for flow-connected pairs, where the subscripts denote the upstream and downstream locations of the pair, and $\pi_{ij} = 0$ otherwise. To form each neighbor set $M(\boldsymbol{s}_i)$, we rank the eligible candidates (earlier observations in the ordering for an observation location, all observations for a prediction location) by placing flow-connected candidates first, each group sorted by stream distance, and keep the first $m$. Since tail-up covariance is zero between flow-unconnected locations, this choice fills the neighbor set with informative neighbors whenever enough flow-connected candidates exist. This step currently evaluates all observation--observation and prediction--observation pairs and is the most computationally expensive part of the S3N workflow; Section \ref{sec:discussion} describes planned improvements.

\subsection{Estimation and prediction with S3Ns}\label{sec:est_pred}

Estimation and prediction for NNGP models can be performed using Bayesian or frequentist methods. The \texttt{spNNGP} R package provides full posterior distributions using MCMC, while the \texttt{BRISC} R package (named after the Bootstrap for Rapid Inference on Spatial Covariances) estimates parameter values using maximum likelihood estimation and estimates their confidence intervals using a bootstrap since MLE asymptotic guarantees are limited in the infilling asymptotic paradigm typical of many spatial problems. \cite{Saha2018_BRISC} demonstrate that \texttt{BRISC} is much faster than \texttt{spNNGP} for both estimation and prediction while providing comparable accuracy. Since our primary goal is scalability, we implement S3N models within a frequentist inferential framework. 
Our estimation and prediction code, released as the \texttt{S3N} R package, is built upon the existing \texttt{BRISC} R package \citep{Saha2018_BRISC}, which uses efficient numerical linear algebra algorithms for matrix and vector computation as described by \cite{Finley2022_spNNGP}. The \texttt{BRISC} package is also implemented primarily in C with calls to the Fortran LAPACK library. All these features contribute to reducing the computational cost of estimation and prediction with the S3N model.

Beyond computational speed, this framework also provides distinct inferential advantages. Confidence intervals for the fixed effect parameters in both S3N and SSN models can be obtained for free during estimation using the estimated variance $\left(X^T \hat{\tilde\Sigma}^{-1} X\right)^{-1}$. However, the underlying BRISC bootstrap method allows for the estimation of confidence intervals for covariance parameters as well, a feature that is not currently available in \texttt{SSN2}. BRISC is a more computationally efficient adaptation of a parametric bootstrap proposed by \cite{Olea2011_bootstrap}, in which the data are decorrelated before sampling and recorrelated afterward by multiplication with either the Cholesky factor or its inverse.

Estimating fish abundance over an entire region based on observed values requires prediction at unobserved locations. Given observations $\boldsymbol{Y_1} = (Y(\boldsymbol{s_1}), ..., Y(\boldsymbol{s_n}))$ at a set of locations $\mathcal{D}_1 = \{\boldsymbol{s_1}, ..., \boldsymbol{s_n}\}$ with covariates $\boldsymbol{X}_1$, we wish to predict the response process $\boldsymbol{Y_2} = (Y(\boldsymbol{r_1}), ..., Y(\boldsymbol{r_k}))$ at another set of locations $\mathcal{D}_2 = \{\boldsymbol{r_1}, ..., \boldsymbol{r_k}\}$ with covariates $\boldsymbol{X}_2$ in order to map fish abundance across the region of interest and estimate regional population sizes. Let $\boldsymbol\Sigma_{11} := \text{Cov}(\boldsymbol{Y_1}, \boldsymbol{Y_1})$, $\boldsymbol\Sigma_{12} := \text{Cov}(\boldsymbol{Y_1}, \boldsymbol{Y_2})$, and so on; then the joint distribution of $\boldsymbol{Y_1}$ and $\boldsymbol{Y_2}$ under a Gaussian process is
\begin{equation}
    \begin{bmatrix}
        \boldsymbol{Y_1} \\
        \boldsymbol{Y_2}
    \end{bmatrix}
    \sim N \left( 
    \begin{bmatrix}
        \boldsymbol{X}_1 \\
        \boldsymbol{X}_2
    \end{bmatrix}\boldsymbol\beta,
    \begin{bmatrix}
        \boldsymbol\Sigma_{11} & \boldsymbol\Sigma_{12} \\
        \boldsymbol\Sigma_{21} & \boldsymbol\Sigma_{22}
    \end{bmatrix}
    \right),
\label{eq:joint-dist-marginal}
\end{equation}
and the conditional distribution of $\boldsymbol{Y_2}$ given $\boldsymbol{Y_1}$, $\boldsymbol{X}_1$ and $\boldsymbol{X}_2$ is
\begin{equation}
    \boldsymbol{Y_2} \mid \boldsymbol{Y_1}, \boldsymbol{X}_1, \boldsymbol{X}_2 \sim N\left(\boldsymbol{M}, \boldsymbol{Q}\right), \text{ where}
    \label{eq-cond-dist-norm1-marginal}
\end{equation}
\begin{equation}
    \boldsymbol{M} = \boldsymbol{X}_2\boldsymbol\beta + \boldsymbol\Sigma_{21}\boldsymbol\Sigma_{11}^{-1}\left(\boldsymbol{Y_1} - \boldsymbol{X}_1\boldsymbol\beta\right), \quad
    \boldsymbol{Q} = \boldsymbol\Sigma_{22} - \boldsymbol\Sigma_{21}\boldsymbol\Sigma_{11}^{-1} \boldsymbol\Sigma_{12}.
    \label{eq-cond-dist-norm2-marginal}
\end{equation}

For the S3N model, as for the NNGP, prediction at a new location $\boldsymbol{r}$ conditions only on its neighbor set $M(\boldsymbol{r})$ among the observations (Section \ref{sec:preprocessing}):
\begin{equation}
    Y(\boldsymbol{r}) \mid \boldsymbol{Y}(M(\boldsymbol{r})) \sim N\left(\boldsymbol{x}(\boldsymbol{r})^T\boldsymbol\beta + \boldsymbol{a}_r^T\left\{\boldsymbol{Y}(M(\boldsymbol{r})) - \boldsymbol{X}(M(\boldsymbol{r}))\boldsymbol\beta\right\},\ d_r\right),
\end{equation}
where $\boldsymbol{a}_r$ and $d_r$ are defined as in Equation \ref{eq-NNGP-model} with $\boldsymbol{s}_i$ replaced by $\boldsymbol{r}$, and $\boldsymbol\beta$ and the covariance parameters are replaced by their estimates. Each prediction requires only an $m \times m$ solve, so predicting at $k$ locations costs $O(km^3)$.

During the preparation of this manuscript, we used Anthropic's Claude Opus 5.5 to assist with language editing, LaTeX formatting, and code auditing. No artificial intelligence tools were used to formulate the statistical model, generate data, or interpret results. All AI-assisted text and outputs were thoroughly reviewed, edited, and verified by the authors, who take full responsibility for the accuracy and integrity of the manuscript.

\section{Simulation studies}\label{sec:benchmarking-validation}

\begin{figure}[t]
    \centering
    \includegraphics[width=\linewidth]{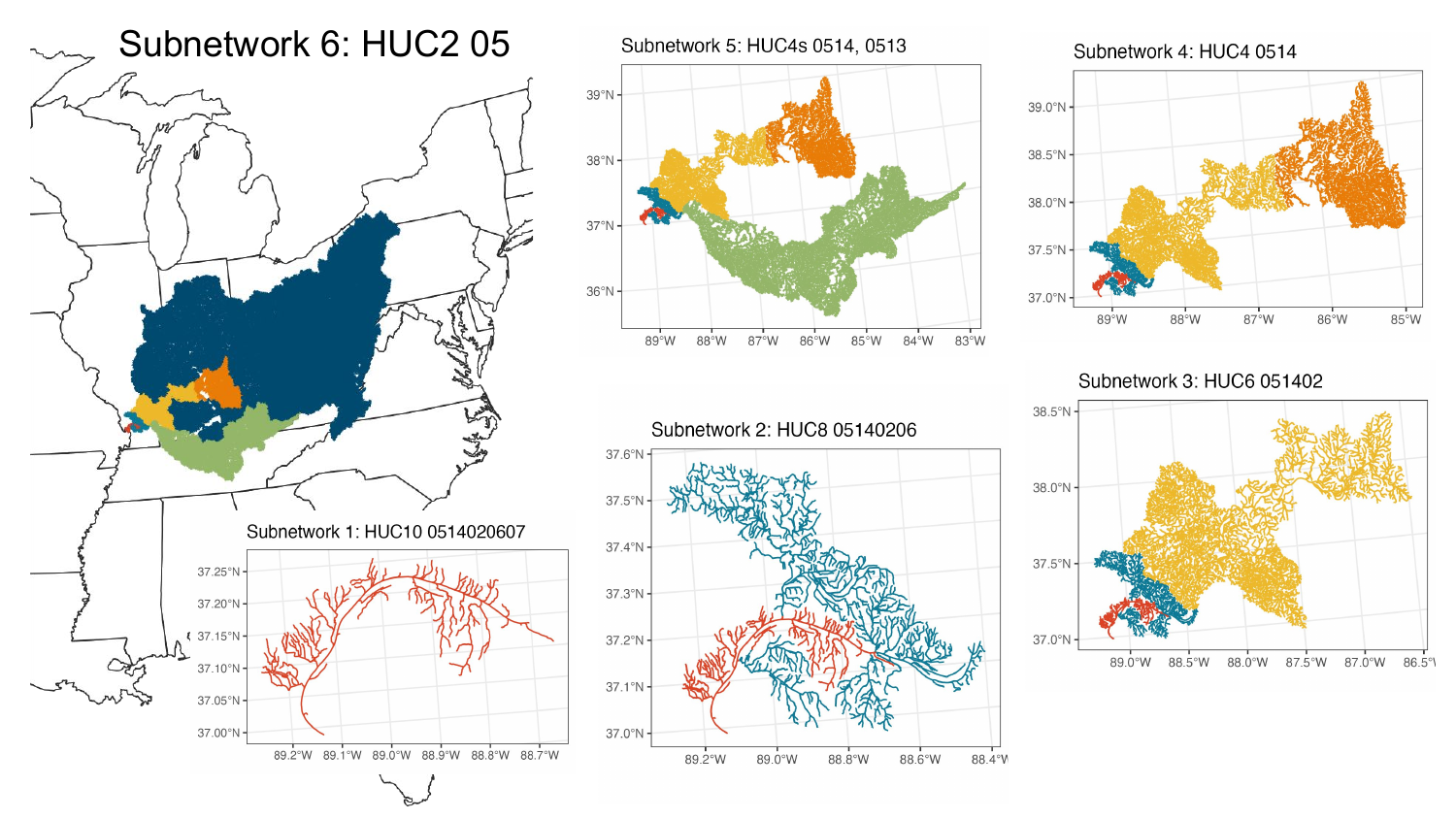}
    \caption{The six subnetworks of the Ohio River Basin used for benchmarking and validation. Network $i$ is a subnetwork of Network $i+1$ for $i = 1, ..., 5$.}
    \label{fig:benchmark-networks}
\end{figure}

We conducted simulation studies to evaluate the accuracy and computational savings of S3Ns. We chose six nested stream networks of increasing size, ranging from 284 to 169,060 stream reaches (Figure \ref{fig:benchmark-networks}). The largest of the six networks is the Ohio River Basin, also known as Region 5 by the United States Geological Survey (USGS). We used the flowline and prediction point shapefiles from the National Stream Internet (NSI), which spans streams across the United States \citep{Nagel2017_NSIHydroData}; these flowlines are high-resolution digital maps of streams that were developed for use with spatial stream network models to ensure that topological assumptions of the spatial stream network models are met. All spatial data are in the NAD 1983 Albers Equal Area Conterminous U.S. projected coordinate system, and all distances are reported in kilometers.

Throughout this section, ``S3N'' refers to our model, preprocessed in R according to Section \ref{sec:preprocessing} (code in the paper repository; see the Data and code availability statement) and fit with the \texttt{S3N} R package (version 1.0.5, GitHub commit 55d9018) using $m = 10$ neighbors and a random ordering of the observation points. ``SSN'' refers to the SSN model preprocessed with \texttt{SSNbler} (version 1.1.1, GitHub commit 0addf3e) and fit with \texttt{SSN2} (version 0.4.0), using \texttt{ssn\_create\_bigdist()} and \texttt{ssn\_lm()} with its big-data approximation (\texttt{local = TRUE}) for all sample sizes. For \texttt{SSNbler}, we ran \texttt{lines\_to\_lsn()} with \texttt{check\_topology = TRUE} and its default \texttt{snap\_tolerance = 0}, so it checked for topological errors but did not snap or otherwise correct any. The \texttt{SSN2} big-data approximation partitions the observations into about $n/200$ groups using k-means clustering on their coordinates, so that groups contain about 200 observations on average, and treats groups as independent during estimation \citep{VerHoef2023_IndexingSPIN, Dumelle2024_SSN2}. It is therefore equivalent to the exact model for $n \le 200$. Because the k-means partition is random, \texttt{SSN2} estimates can differ between runs when $n > 200$; refitting our simulated data sets changed $\hat\beta_0$ by up to about 0.6 standard errors. S3N estimates parameters by maximum likelihood, while SSN2 uses its default, restricted maximum likelihood.

Each simulation replicate ran as a single job with 4 CPU cores on an AMD EPYC node (7H12 or 9684X processors) of Montana State University's Tempest research computing cluster, using R 4.5.2 with \texttt{OMP\_NUM\_THREADS} set to 4. \texttt{SSNbler}'s \texttt{lines\_to\_lsn()} used 4 cores and \texttt{ssn\_create\_bigdist()} a single core, while S3N's estimation code, which is parallelized with OpenMP, and \texttt{SSN2}'s linear algebra, through the multithreaded OpenBLAS library, could each use up to 4 threads. Because S3N and SSN were run back to back within the same job for each replicate, every runtime comparison between the two was made on the same hardware.

For each network, we drew a random sample of the reach midpoints to represent observation locations, and at these locations we simulated responses from the following SSN model with exponential tail-up covariance and two fixed effects, an intercept $\beta_0$ and a single covariate $\beta_1$:
\begin{equation}
        \boldsymbol{Y} = \boldsymbol{X}\boldsymbol\beta + \boldsymbol{w}, \quad \boldsymbol{w} \sim N\left(0, \boldsymbol{\Sigma}\right), \quad
        \Sigma\big(\boldsymbol{s_i}, \boldsymbol{s_j}\big) = \pi_{ij} \sigma^2 \exp(-h_{ij}/\lambda) + \tau^2 \delta_{ij}.
\label{eq-bench-sim-model}
\end{equation}

It is worth emphasizing that we simulated the data from the exact SSN model; both S3N and, for $n > 200$, \texttt{SSN2}'s big-data approximation are approximations to it. Here, $\tau^2$ is the independent variance, $\sigma^2$ is the exponential tail-up spatial variance scaling parameter, and $\lambda$ is the range parameter representing the characteristic distance for spatial correlations. We simulated from this model with intercept $\beta_0 = -44$, elevation coefficient $\beta_1 = 0.5$, $\sigma^2 = 5$, $\tau^2 = 0.1$, and $\lambda = 5$ km, values chosen to resemble estimates from the real fish data in Section \ref{sec:case-study}.

To separately evaluate the impact on runtime of the number of observations and the number of reaches, we conducted two experiments: (1) holding the number of observation points fixed at 1,000 and varying the number of reaches from 1,273 (Network 2) to 169,060 (Network 6), and (2) holding the number of reaches fixed at 169,060 (Network 6) and varying the number of observation points from 100 to 10,000. We found similar results using fewer observation points (200 instead of 1,000) in Experiment 1 and a smaller network (Network 4 instead of Network 6) in Experiment 2, so we present here the results for the larger number of observation points and the larger network. For each combination of river network and set of observation points, we simulated data 50 times from the SSN defined above. Network 1 (284 reaches, 200 observation points) was used for validation only.

\begin{figure}[ht!]
    \centering
    \begin{subfigure}{0.98\linewidth}
        \includegraphics[width=\textwidth]{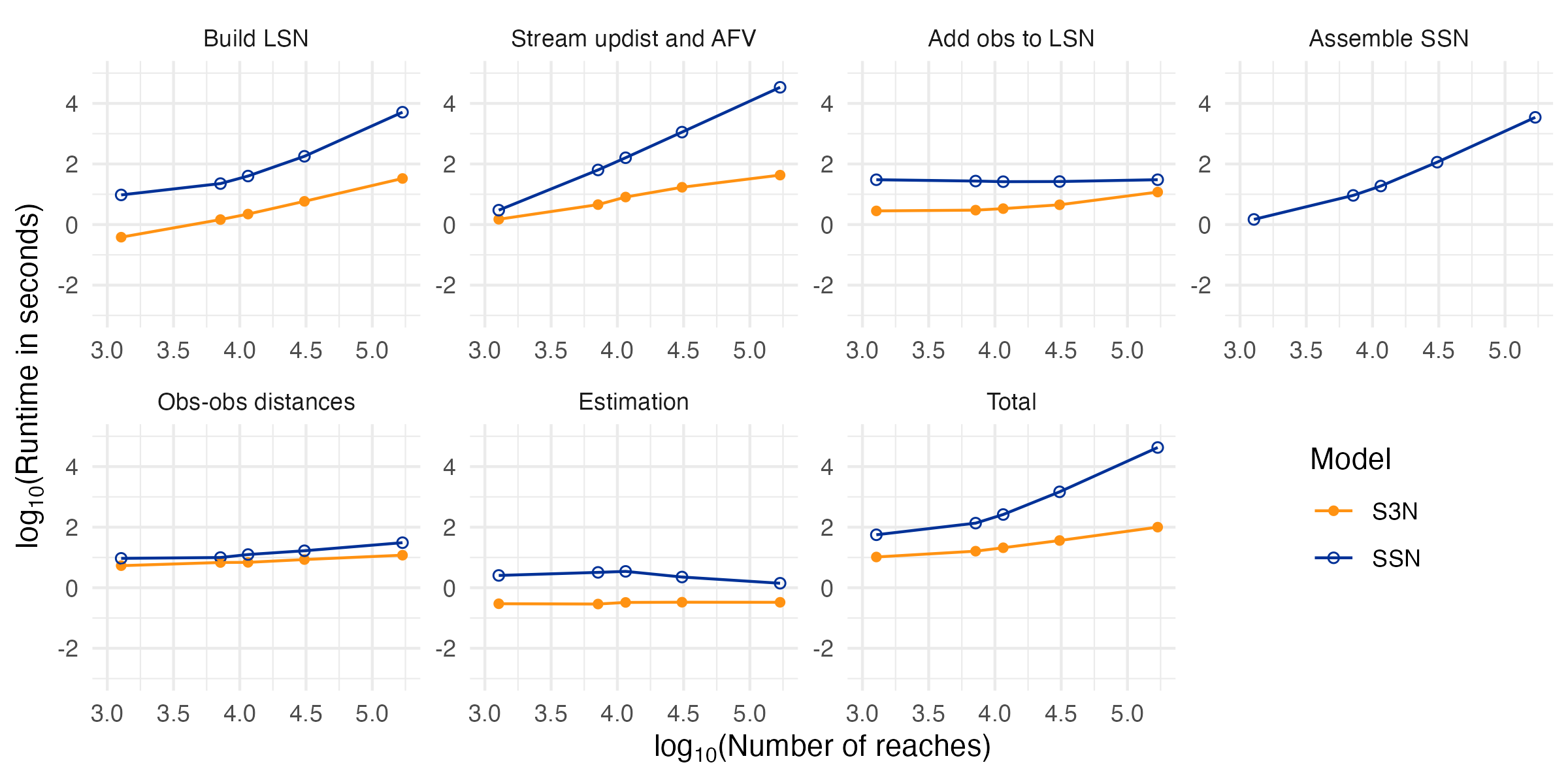}
        \caption{Fixed number of observation locations (1,000), varying number of reaches (1,273-169,060).}
        \label{fig:bench-vary-reach}
    \end{subfigure}
    \hfill
    \begin{subfigure}{0.98\linewidth}
        \centering
        \includegraphics[width=\linewidth]{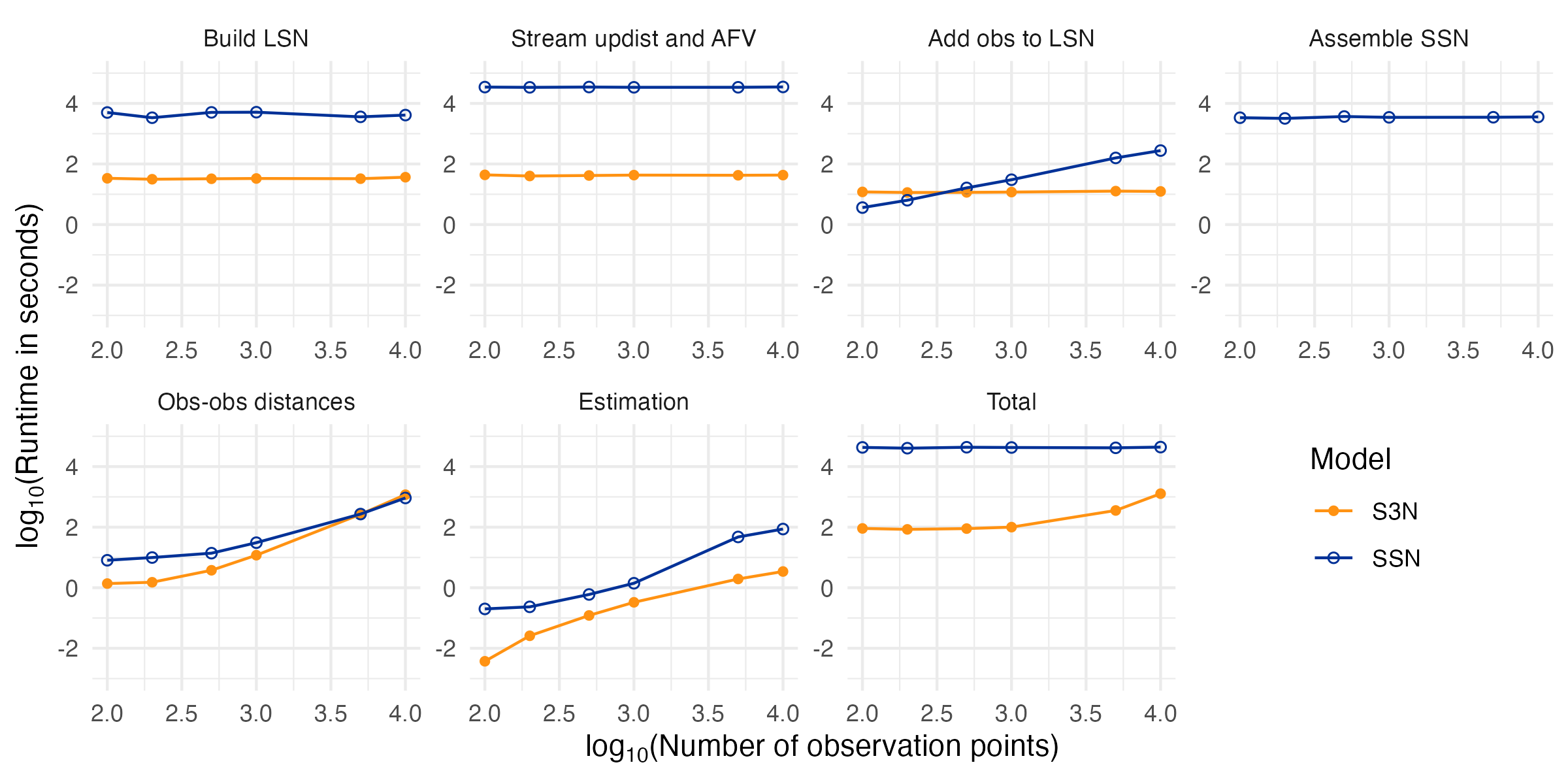}
        \caption{Fixed number of reaches (169,060), varying number of observation locations (100-10,000).}
        \label{fig:bench-vary-nobs}
    \end{subfigure}
    \caption{Benchmarking to compare the computational cost of S3N and SSN as the number of reaches or observation points increases. The first two preprocessing steps, building the stream network and computing stream variables, were up to about 150 and 844 times as fast with S3N as with SSN, respectively. Estimation was as much as 54 times as fast with S3N.}
    \label{fig:bench-results}
\end{figure}

Figure \ref{fig:bench-results} shows the results of the two experiments. Unsurprisingly, for both the S3N and SSN models, the computational time required for building the landscape network and computing stream variables increases with the number of reaches and not with the number of observations. Computational time for these two steps is up to three orders of magnitude faster for S3N than SSN. For building the landscape network, the log-log plot of computational time against the number of reaches is approximately linear for S3N, with a fitted slope of 0.92 (95\% CI 0.79--1.05), indicating roughly linear growth in the number of reaches \citep{Sedgewick2011_algorithms}. For SSN, the local slope increases from 0.5 for the smallest networks to 2.0 for the largest, approaching quadratic growth. Computing stream variables, a step that involves no topology checks in either pipeline, shows the starkest difference: SSN's computational time grows almost exactly quadratically (fitted slope 1.92), whereas S3N's grows sublinearly (fitted slope 0.70), yielding speedups of about 800 times on the largest network.

S3N and \texttt{SSNbler} differ in their topological checks, which are part of building the landscape network. S3N currently checks for complex confluences and downstream divergences (forks), and it does not yet check for unsnapped nodes, dangling nodes, intersections without nodes, or converging nodes. \texttt{SSNbler} currently checks for all of these topological errors, but only corrects unsnapped nodes; any other topological errors have to be manually corrected. The preprocessing in the simulations presented here includes whatever topological checks each pipeline performs. All of the topological errors that S3N does not yet assess can be detected near stream outlets. Because an efficient nearest-neighbor search can find the reach endpoints close to each outlet without comparing every pair, we expect that building the landscape network with S3N will remain approximately linear, whereas our simulations suggest that \texttt{SSNbler}'s local slope approaches 2 as in a quadratic scaling.

The computational cost of adding observation points to the landscape network is constant with the number of observation points for S3N and increases roughly linearly for SSN. The computational time required for assembling the SSN object with \texttt{SSNbler}, a step S3N does not require, grows with the number of reaches, requiring approximately 20 seconds for 11,540 reaches and an hour for 169,060 reaches. Computing the pairwise distances among observation points is comparable between the two models, with S3N slightly slower at 10,000 observation points.

As one would expect, for both models, the computational time required for estimation depends on the number of observations and not on the number of reaches. Estimation was 4--13 times as fast for S3N as for SSN with up to 1,000 observation points (with one outlier of 54 times for the smallest sample on Network 6, where both fits took under a second) and 24--52 times as fast with 5,000--10,000 observation points. For preprocessing and model fitting combined, S3N was about 2.6--2.7 orders of magnitude faster than SSN on the largest network with up to 1,000 observation points. The advantage narrows as the number of observation points grows, because pairwise distance computation, which is comparable for the two models, then dominates both pipelines: with 10,000 observation points, S3N was about 34 times as fast.

\begin{figure}[t]
    \centering
    \includegraphics[width=\linewidth]{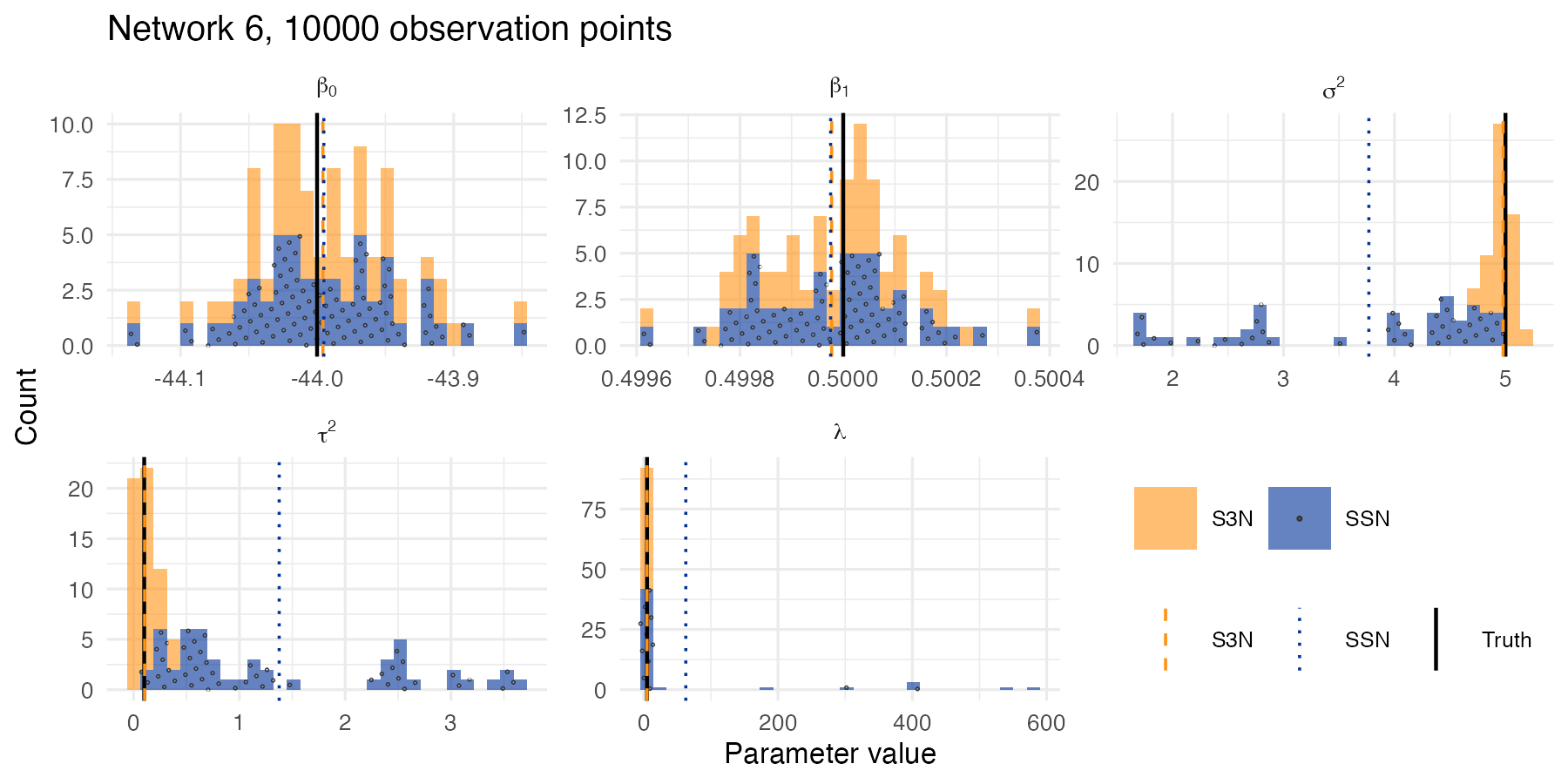}
    \caption{Validation results for Network 6 with 10,000 observation points. S3N parameter estimates are close to the true values. S3N and SSN fixed effects have comparable RMSE, while S3N tends to have lower RMSE for the covariance parameters in this setting. Fixed-effect results were similar across networks and numbers of observation points; covariance-parameter results depended on sampling density.}
    \label{fig:validation-results}
\end{figure}

Figure \ref{fig:validation-results} and Tables \ref{tab:simulation_results_1}-\ref{tab:simulation_results_2} show the validation results. The tables pool replicates across all sets of observation points simulated on each network, so the number of replicates per network ranges from 50 to 300. Recall that these data were simulated from an SSN model. Both models have more difficulty estimating the individual covariance parameters when observations are sparse relative to the size of the network, consistent with the fact that only the ratio $\sigma^2/\lambda$ is consistently estimable under infill asymptotics \citep{Ying1991_asymptotic, Zhang2004_asymptotic}. In these settings, SSN tends to attribute spatial variance to the nugget; for example, on Network 6 with 1,000 observation points, average SSN estimates were $\hat\sigma^2 = 0.46$ and $\hat\tau^2 = 4.70$, compared with 4.57 and 0.53 for S3N (true values 5 and 0.1). This pattern is not an artifact of \texttt{SSN2}'s big-data approximation: with 200 or fewer observation points, where \texttt{SSN2} fits the exact model, SSN estimates were close to the true values on the two smallest networks but showed the same shift of variance to the nugget on larger networks (e.g., Network 4 with 100 observation points: $\hat\sigma^2 = 0.96$ and $\hat\tau^2 = 4.25$ for SSN versus 4.78 and 0.34 for S3N). In the settings with 200 or fewer observation points, where \texttt{SSN2} fits the exact SSN model, the RMSE of S3N's fixed-effect estimates was within 3.5\% of the exact model's (ratios 0.97--1.01), and 95\% confidence interval coverage was 0.94 for both models, indicating that the nearest-neighbor approximation cost little or no fixed-effect accuracy at these sample sizes. Coverage of nominal 95\% confidence intervals ranged from 0.93 to 0.96 for S3N and from 0.94 to 0.96 for SSN, within Monte Carlo error of the nominal level, and for both models the average model-based standard errors closely matched the Monte Carlo standard deviations of the estimates (ratios of 0.84--1.19 for S3N and 0.85--1.21 for SSN across settings).

\begin{table}[tbp]
\centering
\caption{Evaluating estimates of $\beta_0 = -44$ by network. Coverage is for nominal 95\% Wald intervals based on each fit's model-based standard error \citep{Morris2019_coverage}. Rows pool replicates across sets of observation points (50 for Network 1; 100 for Networks 2, 3, and 5; 300 for Networks 4 and 6), so Monte Carlo standard errors for coverage are about 0.031, 0.022, and 0.013, respectively.}
\label{tab:simulation_results_1}
\small
\begin{tabular}{c
  S[table-format=-1.3] S[table-format=-1.3]
  S[table-format=1.2]  S[table-format=1.2]
  S[table-format=1.2]  S[table-format=1.2]}
\toprule
& \multicolumn{2}{c}{Bias} & \multicolumn{2}{c}{RMSE} & \multicolumn{2}{c}{Coverage} \\
\cmidrule(lr){2-3} \cmidrule(lr){4-5} \cmidrule(lr){6-7}
Network & {S3N} & {SSN} & {S3N} & {SSN} & {S3N} & {SSN} \\
\midrule
1 & -0.081 & -0.11  & 1.37 & 1.40 & 0.96 & 0.94 \\
2 &  0.12  &  0.12  & 0.69 & 0.67 & 0.95 & 0.96 \\
3 &  0.055 &  0.061 & 0.79 & 0.79 & 0.94 & 0.95 \\
4 &  0.056 &  0.056 & 0.38 & 0.38 & 0.94 & 0.94 \\
5 & -0.016 & -0.019 & 0.26 & 0.27 & 0.96 & 0.96 \\
6 & -0.010 & -0.011 & 0.21 & 0.21 & 0.96 & 0.96 \\
\bottomrule
\end{tabular}
\end{table}

\begin{table}[tbp]
\centering
\caption{Evaluating estimates of $\beta_1 = 0.5$ by network}
\label{tab:simulation_results_2}
\small
\begin{tabular}{c
  S[table-format=-1.5] S[table-format=-1.5]
  S[table-format=1.5]  S[table-format=1.5]
  S[table-format=1.2]  S[table-format=1.2]}
\toprule
& \multicolumn{2}{c}{Bias} & \multicolumn{2}{c}{RMSE} & \multicolumn{2}{c}{Coverage} \\
\cmidrule(lr){2-3} \cmidrule(lr){4-5} \cmidrule(lr){6-7}
Network & {S3N} & {SSN} & {S3N} & {SSN} & {S3N} & {SSN} \\
\midrule
1 &  0.0012  &  0.0015  & 0.013   & 0.013   & 0.96 & 0.94 \\
2 & -0.00061 & -0.00061 & 0.0055  & 0.0053  & 0.94 & 0.96 \\
3 & -0.00037 & -0.00041 & 0.0059  & 0.0059  & 0.95 & 0.96 \\
4 & -0.00031 & -0.00031 & 0.0023  & 0.0023  & 0.93 & 0.94 \\
5 &  0.00010 &  0.00011 & 0.00095 & 0.00096 & 0.94 & 0.94 \\
6 &  0.00004 &  0.00004 & 0.00056 & 0.00056 & 0.94 & 0.94 \\
\bottomrule
\end{tabular}
\end{table}

\section{Application: Ohio River Basin fish population estimates}\label{sec:case-study}

Having validated the accuracy and efficiency of S3Ns through simulation, we now apply the S3N model to estimate the distributions and total population sizes for 285 species of fish across the entire Ohio River Basin (USGS hydrologic region 05), a dense river network comprising nearly 170,000 reaches within an area of approximately 420,000 km$^2$, using observations at almost 9,000 locations. This is the same stream network used in benchmarking as Network 6. The Ohio River Basin supports a wide variety of ecological niches and self-sustaining fish populations.

\subsection{Fish survey data}

In this study, we analyzed contemporary (1990--2023) fish community assemblage data (i.e., fish species counts) from 8,924 locations across the Ohio River Basin. This dataset is comprised of three national fish observation datasets\textemdash \cite{Chen2023_Shifting_FishScales}, \cite{GiamOlden2016_FishData}, and \cite{Strecker2011_OldenLCRB}\textemdash which themselves compile data from national and regional programs such as the United States Environmental Protection Agency (USEPA) Environmental Monitoring and Assessment Program (EMAP), the United States Geological Survey (USGS) National Water Quality Assessment Program (NAWQA), numerous state agency biomonitoring programs, and large-scale sampling efforts by other organizations and universities. Together, these datasets provide the most comprehensive observations currently available, offering fish data that cover a large spatial extent but retain the fine spatial resolution necessary to evaluate fish species abundance at ecologically relevant scales.

To ensure comparability, we included only surveys that used standardized backpack or boat electrofishing protocols designed to characterize the entire stream fish community \citep{Bonar2009_standard} and to monitor stream fish uniformly across the US \citep{Barbour1999_rapid, Moulton2002_revised}. Each survey sampled a defined stream length sufficient to characterize local community structure, and all captured individuals were identified to species and counted. We excluded records with unknown gear, which were likely duplicates, and retained only the most recent post-1990 observation of each species on each stream reach. Scientific names were harmonized to FishBase using the \texttt{rfishbase} R package \citep{Boettiger2012_rfishbase}.

One challenge in modeling fish populations across large regions is the range in stream sizes. \cite{Isaak2017_SSN} used densities of fish per 100-m stream length instead of fish per water volume or per cross-sectional river area because they focus on larger fish in smaller streams. To account for this in our study, we assume the length of stream sampled within each survey was 20 times the mean bankfull channel width of the stream reach in which the sampling took place. We assumed a minimum and maximum stream length of 100 m and 1000 m, respectively, replacing the lengths of any sampling event that fell outside of this range. This is consistent with federal and most state fish monitoring electrofishing protocols \citep{HauerLamberti2006_methods}. For the purposes of this study, the above protocol allows us to express the observed counts as densities by dividing each count by its corresponding estimated sampling length.

\subsection{Environmental covariate data}

We use nine environmental covariates to capture descriptions of watershed topography, land use, hydrology, and temperature, all key drivers of fish abundance \citep{Jackson2001}. Covariates include mean elevation (m); upstream watershed area (km$^2$); developed land cover (percentage of land with low, medium, or high development); agricultural land cover (percentage land used for pasture/hay or crops); total annual runoff (m$^3$ per year); mean annual air temperature (degrees Celsius); baseflow index; hydrological alteration index; and floodplain integrity. We use hydrological alteration index data and floodplain integrity data from \cite{McManamay2022_HAI} and \cite{Morrison2023_FloodplainDegradation}, respectively, and all other variables are from the National Hydrography Dataset (NHD) Plus Version 2 Dataset \citep[NHDPlus V2 data, model version 2.1,][]{Wieczorek2018_covariateData}. In our region of interest, there were 806 missing values for hydrological alteration index and 299 for floodplain integrity, representing 0.48\% and 0.18\% of the stream network. We imputed these values by using the value from the nearest stream reach whose value was not missing.

\subsection{Spatial model}

Given observed fish densities $\boldsymbol{Y} = (Y(\boldsymbol{s_1}), ..., Y(\boldsymbol{s_n}))$ in fish per 100-m stream length at $n$ point locations $\boldsymbol{s_1}$, ..., $\boldsymbol{s_n}$ across a region, we wish to estimate the geographic distribution of fish across the region and make regional population estimates by species. We fit the S3N model (Equation \ref{eq-S3N-def}) with the nine covariates described above and an exponential tail-up covariance with no tail-down or Euclidean components,
\begin{equation}
    \Sigma_s\big(\boldsymbol{s_i}, \boldsymbol{s_j}\big) = \pi_{ij}\, \sigma_u^2 \exp\big(-h(\boldsymbol{s_i}, \boldsymbol{s_j})/\lambda_u\big) + \tau^2\delta_{ij}.
    \label{eq-fish-est-S3N-model}
\end{equation}

We use a tail-up covariance because neighbor selection for models that combine tail-up and tail-down components is the subject of ongoing work (see Section \ref{sec:discussion}). Tail-up covariance captures dependence among flow-connected locations, but it does not capture dependence between flow-unconnected locations that may arise from fish moving upstream into different tributaries, so the parameter estimates for such species should be interpreted with this limitation in mind. For this demonstration, we use $m=10$ neighbors, and the ordering of the observation points is simply the row order of the fish data.

We assume a normal distribution to model fish densities, an approach which we suggest is adequate for demonstration purposes in this species-rich region, especially since we are ultimately interested in aggregating results over large regions. While this choice admits the possibility of negative predicted fish densities, we demonstrate in the next section that negative densities are exceedingly rare and small in magnitude, thus likely to correspond to locations where a species is absent or occurs at very low density. Before computing regional population estimates, we set negative densities to zero, so negative predicted densities are interpreted as an indication from the model that no or few fish are expected to be present in that location. Section \ref{sec:discussion} discusses potential future directions to extend this model to generalized linear models.

Preprocessing, including the pairwise distance calculations, depends only on the stream network and the observation and prediction locations, so we performed them once and shared the results across all 285 species; estimation, inference, and prediction were repeated for each species. We configured the stream network twice: once to check its topology and identify network components, and again after removing one complex confluence and restricting the network to its largest connected component, which retained 99.78\% of the network. We then computed stream variables, stream distances, and neighbor sets as described in Section \ref{sec:preprocessing}. Because distances for different prediction locations can be computed independently, we computed the prediction--observation distances for the 160,170 prediction locations without observations in 33 parallel batches of at most 5,000 each; for the 8,924 prediction locations at observed reaches, we reused the observation--observation distances.

For each species, we estimated model parameters using maximum likelihood estimation and estimated confidence intervals for both fixed effects and covariance parameters. To produce regional population size estimates for each species, we computed the predictive mean fish density per 100-m stream length at the midpoint of each stream reach, multiplied this by the length of the stream reach and by appropriate scale units to obtain an estimate for the mean number of fish in each stream reach, then summed across the reaches of the network to obtain a regional population estimate for that species. We use the estimated environmental fixed effects to examine the relationship between these variables and fish abundance.

\subsection{Computational cost}

We ran the entire analysis on a laptop (MacBook Pro 2020, 1.4 GHz Quad-Core Intel Core i5, 8 GB memory) except the pairwise distance computations, which we ran on a research cluster. Network configuration, including identifying and removing the complex confluence and reconfiguring the network, took 3.8 minutes in total. Pairwise distance computation, the rate-limiting part of preprocessing, took 20 minutes for observation pairs and 26 minutes for prediction--observation pairs in 33 parallel batches.

Estimating model parameters and predicting fish densities across Region 5 took 5.5 and 3.3 seconds on average per species. Inference with the bootstrap took approximately as long as estimation for each bootstrap replication (4.9 seconds). The total runtimes for estimation, inference and prediction for all 285 species with 20 bootstrap reps per species were 26 minutes, 466 minutes (7.8 hours), and 16 minutes, respectively.  If confidence intervals for the covariance parameters are not of interest, only for the fixed effects, then the longest step\textemdash the bootstrap\textemdash can be omitted. Without the bootstrap, the total time for preprocessing, estimation, and prediction was 1 hour 32 minutes for all 285 species, approximately 50 minutes for a single species.

\subsection{Results}

\begin{figure}[t!]
    \centering
    \includegraphics[width=\textwidth]{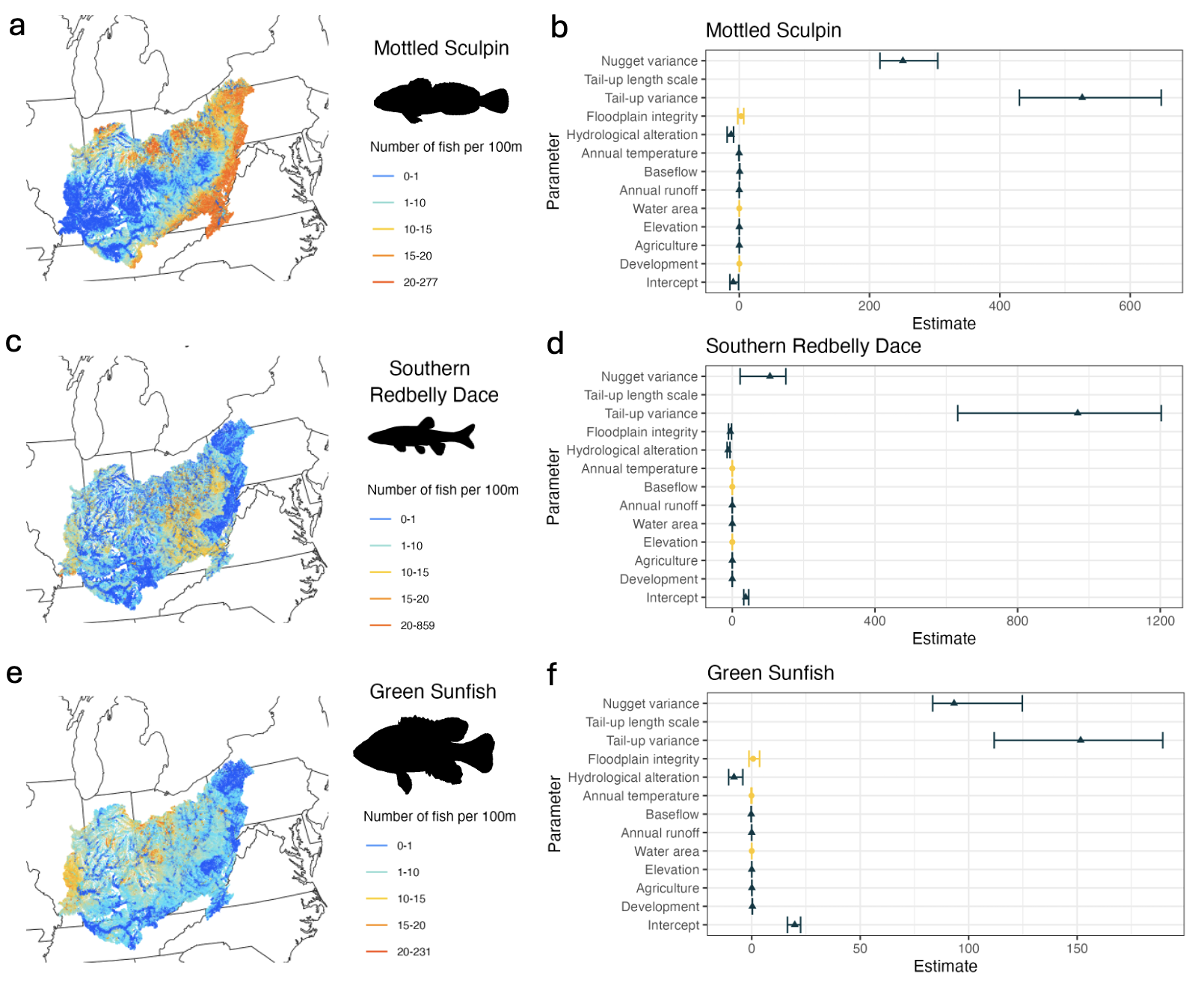}
    \caption{Maps of predictive mean fish density (fish per 100-m stream length), and estimates and confidence intervals of fixed effects and covariance parameters for three example species. Statistically significant coefficients are shown in dark blue with triangles for point estimates, while other coefficients are shown in yellow with circles for point estimates.}
    \label{fig:maps-params-ORB}
\end{figure}

In-sample species-level correlations between predicted and observed densities across sites ranged from 0.52 to 1.0 (mean = 0.96, SD = 0.08), while community-level correlations across species within each site averaged 0.80 (SD = 0.22), ranging from $-0.29$ to 1.0. The ratio of predicted to observed counts ranged from 0.2 to 1.02 on reaches where the species in question was observed, and from 0.57 to 3.23 on reaches where any fish were observed.

The spatial distributions of predicted fish density for three example species\textemdash mottled sculpin (\textit{Cottus bairdii}), Southern redbelly dace (\textit{Chrosomus erythrogaster}), and green sunfish (\textit{Lepomis cyanellus})\textemdash correspond with known distributions and dominant environmental drivers \citep{PageBurr1991_fieldguide}. Mottled sculpin in the Ohio River Basin is patchily distributed, occurring primarily in cool, clear, well-oxygenated streams of states such as Ohio, West Virginia, Pennsylvania, and Kentucky (Figure \ref{fig:maps-params-ORB}a), where it inhabits riffles and runs in upland streams with coarse substrates, while avoiding large, warm, or impacted rivers. This is supported by modeled densities that are negatively associated with annual temperature, annual runoff, and hydrologic alteration, and positively associated with elevation (Figure \ref{fig:maps-params-ORB}b). The Southern redbelly dace occurs primarily in small spring-fed or groundwater-influenced headwater streams that are cool, clear, shaded, and have gravel or sand substrates and abundant vegetation (Figure \ref{fig:maps-params-ORB}c). Dace populations have been extirpated in parts of their range where excessive siltation and turbidity are caused by the loss of forested lands, channelization and loss of lateral connectivity to the floodplain, and construction of dams that simplify river habitats; this is consistent with the negative associations of density with developed and agricultural land and hydrological alteration (Figure \ref{fig:maps-params-ORB}d). The green sunfish is widespread throughout the basin (Figure \ref{fig:maps-params-ORB}e), occurring in a variety of warm, slow-flowing streams, reservoirs, and backwaters, and often thriving in disturbed or degraded habitats, as shown by modeled positive associations with developed and agricultural land (Figure \ref{fig:maps-params-ORB}f). These species examples suggest that the S3N model captures a range of geographic distributions and environmental determinants of freshwater fishes in the Ohio River Basin.

We note a small number of instances where the model predicted negative fish densities as a result of using a Gaussian process in the model. However, the mean proportion of observed reaches predicted to have negative densities with magnitude 1 fish/100m or more was just 0.0017 (or 0.17\%) across species. Moreover, predicted negative densities are considerably smaller in magnitude when compared to predicted positive densities. Therefore, the small proportion of predicted negative densities appears to correspond to locations and species for which the model predicts densities near zero.

\section{Discussion}\label{sec:discussion}

This paper introduces a new class of Scalable Spatial Stream Network (S3N) models, which extend nearest-neighbor Gaussian processes to incorporate ecologically relevant spatial dependence while greatly improving computational efficiency over existing spatial stream network models. Using data simulated from an SSN on real river networks, we demonstrate that fixed effect estimates for S3N and SSN have comparable and low RMSE, while covariance parameter estimate performance depended on sampling density. These simulations also demonstrate that the first stream preprocessing steps scale more favorably with network size and are up to three orders of magnitude faster for S3N than SSN for the range we tested in the number of reaches and observation points. Estimation is up to 54 times as fast for S3N as for SSN, and for larger numbers of reaches, the entire sequence of preprocessing and model fitting was up to 2.6--2.7 orders of magnitude faster for S3N than for SSN on the largest network. The estimates presented here for approximately 170,000 reaches and 9,000 observation points in the Ohio River Basin offer a critical step toward mapping the geographic distribution of freshwater fish species and understanding the role of environmental drivers at large scales.

Ongoing work includes implementing tail-down and Euclidean covariance components, as these require additional choices when used with the nearest neighbors approximation, and these choices warrant dedicated evaluation. Computing tail-down covariance for two flow-unconnected points requires the distance from each point to their nearest common junction. Including both stream and Euclidean covariances in an S3N model should ideally use both the nearest neighbors with respect to Euclidean distance, $M_e(\boldsymbol{s_i})$, and those with respect to stream distance, $M_s(\boldsymbol{s_i})$. Let $M(\boldsymbol{s_i}) = M_e(\boldsymbol{s_i}) \cup M_s(\boldsymbol{s_i})$; one approach is to choose the $m_e$ nearest neighbors with respect to Euclidean distance and either the $m_s$ nearest neighbors with respect to stream distance or all observed locations that fall within some radius $r_s$ with respect to stream distance. While covariates likely account for most Euclidean correlation and tail-up covariance captures stream correlations due to downstream flow, including tail-down and Euclidean covariance will provide additional flexibility. 

Another direction for future work is to extend S3Ns to non-Gaussian distributions such as binomial, Poisson, zero-inflated Poisson, and negative binomial. This would better adapt S3Ns for modeling ecological count data. NNGPs have been extended to binary data \citep{Saha2022_probitNNGP, Finley2022_spNNGP}, and Vecchia approximations more generally to non-Gaussian likelihoods \citep{ZilberKatzfuss2021_VecchiaLaplace}, suggesting a path toward S3N models for counts. The \texttt{FRK} R package for fixed-rank kriging, another approach to scalable spatial processes, only added support for non-Gaussian distributions in 2024 \citep{Sainsbury2024_FRK}, although the underlying spatial random effects model had been extended to the exponential family earlier \citep{SenguptaCressie2013_expfam}.

We are also investigating the performance of different orderings of the point locations in defining the nearest neighbors. Under the full model without the nearest-neighbor approximation, the ordering does not matter. However, under the nearest-neighbor approximation, nearest neighbors are chosen only among the earlier points in the ordering, so the quality of the nearest-neighbor approximation depends on the index assigned to the locations \citep{Guinness2018_Permutation}. Future work should compare a few methods empirically and also explore theoretical arguments for using one method versus another.

Currently the slowest step in the S3N implementation is computing the pairwise distances among observation points for estimation and prediction, and between prediction and observation points for prediction. This computational cost can be greatly reduced by using efficient search algorithms and implementing the code in C++, work which is currently underway.

Thus far, we have applied the nearest-neighbor approximation to the response process rather than the latent process. This functionality could easily be added and is desirable if the independent variance is thought to arise solely from measurement error. While the computational savings over the exact SSN model will not be as great for the latent process as for the response process, this option would enable users to make predictions at the latent level instead of the response level if they prefer.

In conclusion, this study introduces a scalable spatial stream network (S3N) model that enables spatial process modeling on stream networks with many reaches and observation points. By extending nearest-neighbor Gaussian processes to a stream-based context and implementing key preprocessing steps more efficiently, we provide a framework capable of handling a computationally demanding task that limits the application of existing approaches. We demonstrate the utility of this approach through large-scale mapping of freshwater fish distributions and quantifying the influence of environmental drivers across broad networks. By combining a valid nearest-neighbor process with flow-aware neighbor selection and efficient network preprocessing, S3Ns make spatial process models practical for stream networks with hundreds of thousands of reaches and thousands of observations, a scale at which existing approaches become prohibitively slow.

\section{Acknowledgments}\label{sec:acknowledgments}

Simulations were performed on the Tempest High Performance Computing System, operated and supported by University Information Technology Research Cyberinfrastructure (RRID:SCR\_026229) at Montana State University. We thank the editor and two anonymous reviewers of an earlier version of this manuscript for comments that substantially improved it.

\section{Funding}\label{sec:funding}

The authors gratefully acknowledge the following source of funding: J.D.O. was supported by the Richard C. and Lois M. Worthington Endowed Professor in Fisheries Management from the School of Aquatic and Fishery Sciences, University of Washington.

\section{Conflicts of interest}\label{sec:disclosure-statement}

The authors have no conflicts of interest to declare.

\section{Data and code availability statement}\label{sec:data-availability-statement}

Hydrological alteration index data and floodplain integrity data are publicly available from \cite{McManamay2022_HAI} and \cite{Morrison2023_FloodplainDegradation}, respectively, and all other covariate data is available from the National Hydrography Dataset (NHD) Plus Version 2 Dataset \citep[NHDPlus V2 data, model version 2.1,][]{Wieczorek2018_covariateData}. The flowline and prediction point shapefiles from the National Stream Internet (NSI) are publicly available \citep{Nagel2017_NSIHydroData}. Fish data were compiled from U.S. federal and state biomonitoring programs and are subject to the data-sharing policies of the respective agencies. As these datasets are owned by the contributing agencies, the original raw data cannot be redistributed by the authors. However, portions of the data needed to reproduce key results can be made available upon reasonable request to Julian Olden. The simulation study uses only the publicly available NSI flowlines and simulated responses, so all simulation results can be fully reproduced from the archived code.

\if1\anon
{
Estimation and prediction for S3N models are implemented in the \texttt{S3N} R package, available at \url{https://github.com/jpierkunke/S3N}. The preprocessing functions described in Section \ref{sec:preprocessing}, along with the code to reproduce all results in this paper, are available at \url{https://github.com/jpierkunke/S3N_paper}. The preprocessing code may be incorporated into the \texttt{S3N} package in a future release.
} \fi

\if0\anon
{
Estimation and prediction for S3N models are implemented in the \texttt{S3N} R package, available at \textit{<GitHub link is redacted during review to hide author identities>}. The preprocessing functions described in Section \ref{sec:preprocessing}, along with the code to reproduce all results in this paper, are available at \textit{<GitHub link is redacted during review to hide author identities>}. The preprocessing code may be incorporated into the \texttt{S3N} package in a future release.
} \fi

\spacingset{1}
\bibliography{bibliography.bib}

@Article{Altermatt2013,
    author={Altermatt, Florian},
    title={Diversity in riverine metacommunities: a network perspective},
    journal={Aquatic Ecology},
    year={2013},
    month={Sep},
    volume={47},
    number={3},
    pages={365-377},
    doi={10.1007/s10452-013-9450-3},
    language={en}
}

@book{Barbour1999_rapid,
  title={Rapid bioassessment protocols for use in wadeable streams and rivers: periphyton, benthic macroinvertebrates and fish},
  author={Barbour, Michael T and Jeroen Gerritsen and Blaine D. Snyder and James B. Stribling},
  year={1999},
  publisher={US Environmental Protection Agency, Office of Water}
}

@article{Bilton2001_dispersal,
  title={Dispersal in freshwater invertebrates},
  author={Bilton, David T and Freeland, Joanna R and Okamura, Beth},
  journal={Annual review of ecology and systematics},
  volume={32},
  number={1},
  pages={159--181},
  year={2001},
  publisher={Annual Reviews 4139 El Camino Way, PO Box 10139, Palo Alto, CA 94303-0139, USA}
}

@article{Boettiger2012_rfishbase,
  title={{rfishbase}: exploring, manipulating and visualizing FishBase data from {R}},
  author={Boettiger, Carl and Lang, D Temple and Wainwright, PC},
  journal={Journal of fish biology},
  volume={81},
  number={6},
  pages={2030--2039},
  year={2012},
  publisher={Wiley Online Library}
}

@article{Bonar2009_standard,
  title={Standard methods for sampling {North American} freshwater fishes},
  author={Bonar, Scott A and Hubert, Wayne A and Willis, David W},
  journal={Standard methods for sampling North American freshwater fishes},
  year={2009}
}

@Article{Bond2011,
    author={Bond, Nick
    and Thomson, Jim
    and Reich, Paul
    and Stein, Janet},
    title={Using species distribution models to infer potential climate change-induced range shifts of freshwater fish in south-eastern {Australia}},
    journal={Marine and Freshwater Research},
    year={2011},
    month={2025/08/04/20:28:33},
    volume={62},
    number={9},
    pages={1043},
    issn={1323-1650},
    doi={10.1071/MF10286},
    language={en}
}

@Article{CampbellGrant2007,
    author={Campbell Grant, Evan H.
    and Lowe, Winsor H.
    and Fagan, William F.},
    title={Living in the branches: population dynamics and ecological processes in dendritic networks},
    journal={Ecology Letters},
    year={2007},
    month={Feb},
    volume={10},
    number={2},
    pages={165-175},
    doi={10.1111/j.1461-0248.2006.01007.x},
    language={en}
}

@article{Chen2023_Shifting_FishScales,
    title={Shifting taxonomic and functional community composition of rivers under land use change},
    author={Chen, Kai and Midway, Stephen R and Peoples, Brandon K and Wang, Beixin and Olden, Julian D},
    journal={Ecology},
    volume={104},
    number={11},
    pages={e4155},
    year={2023},
    publisher={Wiley Online Library}
}

@article{Comte2018_fish,
  title={Fish dispersal in flowing waters: A synthesis of movement-and genetic-based studies},
  author={Comte, Lise and Olden, Julian D},
  journal={Fish and Fisheries},
  volume={19},
  number={6},
  pages={1063--1077},
  year={2018},
  publisher={Wiley Online Library}
}

@book{CressieWikle2011_SpatiotemporalData,
    title={Statistics for spatio-temporal data},
    author={Cressie, Noel and Wikle, Christopher K},
    year={2011},
    publisher={John Wiley \& Sons}
}

@article{Darwall2018,
    author = {Darwall, William and Bremerich, Vanessa and De Wever, Aaike and Dell, Anthony I. and Freyhof, Jörg and Gessner, Mark O. and Grossart, Hans-Peter and Harrison, Ian and Irvine, Ken and Jähnig, Sonja C. and Jeschke, Jonathan M. and Lee, Jessica J. and Lu, Cai and Lewandowska, Aleksandra M. and Monaghan, Michael T. and Nejstgaard, Jens C. and Patricio, Harmony and Schmidt-Kloiber, Astrid and Stuart, Simon N. and Thieme, Michele and Tockner, Klement and Turak, Eren and Weyl, Olaf},
    title = {The Alliance for Freshwater Life: A global call to unite efforts for freshwater biodiversity science and conservation},
    journal = {Aquatic Conservation: Marine and Freshwater Ecosystems},
    volume = {28},
    number = {4},
    pages = {1015-1022},
    doi = {https://doi.org/10.1002/aqc.2958},
    year = {2018}
}

@article{Datta2016_NNGP,
    title={Hierarchical nearest-neighbor {Gaussian} process models for large geostatistical datasets},
    author={Datta, Abhirup and Banerjee, Sudipto and Finley, Andrew O and Gelfand, Alan E},
    journal={Journal of the American Statistical Association},
    volume={111},
    number={514},
    pages={800--812},
    year={2016},
    publisher={Taylor \& Francis}
}

@article{Datta2016_NNGP_Interpretation,
    title={On nearest-neighbor {Gaussian} process models for massive spatial data},
    author={Datta, Abhirup and Banerjee, Sudipto and Finley, Andrew O and Gelfand, Alan E},
    journal={Wiley Interdisciplinary Reviews: Computational Statistics},
    volume={8},
    number={5},
    pages={162--171},
    year={2016},
    publisher={Wiley Online Library}
}

@article{Dent1999_StreamWaterNutrient,
    title={Spatial heterogeneity of stream water nutrient concentrations over successional time},
    author={Dent, C Lisa and Grimm, Nancy B},
    journal={Ecology},
    volume={80},
    number={7},
    pages={2283--2298},
    year={1999},
    publisher={Wiley Online Library}
}

@article{Dumelle2024_SSN2,
    title = {{SSN2}: The next generation of spatial stream network modeling in {R}},
    author = {Michael Dumelle and Erin E. Peterson and Jay M. {Ver Hoef} and Alan Pearse and Daniel J. Isaak},
    journal = {Journal of Open Source Software},
    year = {2024},
    volume = {9},
    number = {99},
    pages = {6389},
    doi = {10.21105/joss.06389},
    publisher = {The Open Journal},
  }

@article{Eros2012characterizing,
  title={Characterizing connectivity relationships in freshwaters using patch-based graphs},
  author={Er{\H{o}}s, Tibor and Olden, Julian D and Schick, Robert S and Schmera, D{\'e}nes and Fortin, Marie-Jos{\'e}e},
  journal={Landscape ecology},
  volume={27},
  number={2},
  pages={303--317},
  year={2012},
  publisher={Springer}
}

@Article{Fausch2002,
    author={Fausch, Kurt D.
    and Torgersen, Christian E.
    and Baxter, Colden V.
    and Li, Hiram W.},
    title={Landscapes to Riverscapes: Bridging the Gap between Research and Conservation of Stream Fishes},
    journal={BioScience},
    year={2002},
    month={2025/08/04/20:30:01},
    volume={52},
    number={6},
    pages={483},
    issn={0006-3568},
    doi={10.1641/0006-3568(2002)052[0483:LTRBTG]2.0.CO;2},
    language={en}
}

@article{Finley2022_spNNGP,
    title={{spNNGP R} package for nearest neighbor {Gaussian} process models},
    author={Finley, Andrew O and Datta, Abhirup and Banerjee, Sudipto},
    journal={Journal of Statistical Software},
    volume={103},
    pages={1--40},
    year={2022}
}

@article{Ganio2005_GeostatApproach,
    title={A geostatistical approach for describing spatial pattern in stream networks},
    author={Ganio, Lisa M and Torgersen, Christian E and Gresswell, Robert E},
    journal={Frontiers in Ecology and the Environment},
    volume={3},
    number={3},
    pages={138--144},
    year={2005},
    publisher={Wiley Online Library}
}

@article{GiamOlden2016_FishData,
    title={Environment and predation govern fish community assembly in temperate streams},
    author={Giam, Xingli and Olden, Julian D},
    journal={Global Ecology and Biogeography},
    volume={25},
    number={10},
    pages={1194--1205},
    year={2016},
    publisher={Wiley Online Library}
}

@article{Guinness2018_Permutation,
    title={Permutation and grouping methods for sharpening {Gaussian} process approximations},
    author={Guinness, Joseph},
    journal={Technometrics},
    volume={60},
    number={4},
    pages={415--429},
    year={2018},
    publisher={Taylor \& Francis}
}

@Article{Harper2021,
    author={Harper, Meagan
    and Mejbel, Hebah S.
    and Longert, Dylan
    and Abell, Robin
    and Beard, T. Douglas
    and Bennett, Joseph R.
    and Carlson, Stephanie M.
    and Darwall, William
    and Dell, Anthony
    and Domisch, Sami
    and Dudgeon, David
    and Freyhof, J{\"o}rg
    and Harrison, Ian
    and Hughes, Kathy A.
    and J{\"a}hnig, Sonja C.
    and Jeschke, Jonathan M.
    and Lansdown, Richard
    and Lintermans, Mark
    and Lynch, Abigail J.
    and Meredith, Helen M. R.
    and Molur, Sanjay
    and Olden, Julian D.
    and Ormerod, Steve J.
    and Patricio, Harmony
    and Reid, Andrea J.
    and Schmidt‐Kloiber, Astrid
    and Thieme, Michele
    and Tickner, David
    and Turak, Eren
    and Weyl, Olaf L. F.
    and Cooke, Steven J.},
    title={Twenty‐five essential research questions to inform the protection and restoration of freshwater biodiversity},
    journal={Aquatic Conservation: Marine and Freshwater Ecosystems},
    year={2021},
    month={Sep},
    volume={31},
    number={9},
    pages={2632-2653},
    doi={10.1002/aqc.3634},
    language={en}
}

@book{HauerLamberti2006_methods,
    title={Methods in stream ecology},
    author={Hauer, F Richard and Lamberti, Gary},
    year={2006},
    edition={2nd},
    publisher={Academic press}
}

@article{Isaak2017_SSN,
    author = {Isaak, Daniel J. and Ver Hoef, Jay M. and Peterson, Erin E. and Horan, Dona L. and Nagel, David E.},
    title = {Scalable population estimates using spatial-stream-network ({SSN}) models, fish density surveys, and national geospatial database frameworks for streams},
    journal = {Canadian Journal of Fisheries and Aquatic Sciences},
    volume = {74},
    number = {2},
    pages = {147-156},
    year = {2017},
    doi = {10.1139/cjfas-2016-0247}
}

@Article{Jackson2001,
    author={Jackson, Donald A.
    and Peres-Neto, Pedro R.
    and Olden, Julian D.},
    title={What controls who is where in freshwater fish communities -- the roles of biotic, abiotic, and spatial factors},
    journal={Canadian Journal of Fisheries and Aquatic Sciences},
    year={2001},
    month={2025/08/04/20:31:58},
    volume={58},
    number={1},
    pages={157-170},
    doi={10.1139/cjfas-58-1-157},
    language={en}
}

@article{KangKatzfuss2023_correlation,
  title={Correlation-based sparse inverse {Cholesky} factorization for fast {Gaussian-process} inference},
  author={Kang, Myeongjong and Katzfuss, Matthias},
  journal={Statistics and Computing},
  volume={33},
  number={3},
  pages={56},
  year={2023},
  publisher={Springer}
}

@article{Leibowitz2018,
    author = {Leibowitz, Scott G. and Wigington Jr., Parker J. and Schofield, Kate A. and Alexander, Laurie C. and Vanderhoof, Melanie K. and Golden, Heather E.},
    title = {Connectivity of Streams and Wetlands to Downstream Waters: An Integrated Systems Framework},
    journal = {JAWRA Journal of the American Water Resources Association},
    volume = {54},
    number = {2},
    pages = {298-322},
    doi = {https://doi.org/10.1111/1752-1688.12631},
    year = {2018}
}

@Article{Lynch2023,
    author={Lynch, Abigail J.
    and Cooke, Steven J.
    and Arthington, Angela H.
    and Baigun, Claudio
    and Bossenbroek, Lisa
    and Dickens, Chris
    and Harrison, Ian
    and Kimirei, Ismael
    and Langhans, Simone D.
    and Murchie, Karen J.
    and Olden, Julian D.
    and Ormerod, Steve J.
    and Owuor, Margaret
    and Raghavan, Rajeev
    and Samways, Michael J.
    and Schinegger, Rafaela
    and Sharma, Subodh
    and Tachamo‐Shah, Ram‐Devi
    and Tickner, David
    and Tweddle, Denis
    and Young, Nathan
    and J{\"a}hnig, Sonja C.},
    title={People need freshwater biodiversity},
    journal={WIREs Water},
    year={2023},
    month={May},
    volume={10},
    number={3},
    pages={e1633},
    doi={10.1002/wat2.1633},
    language={en}
}

@Article{Markovic2012,
    author={Markovic, Danijela
    and Freyhof, J{\"o}rg
    and Wolter, Christian},
    title={Where Are All the Fish: Potential of Biogeographical Maps to Project Current and Future Distribution Patterns of Freshwater Species},
    journal={PLoS ONE},
    year={2012},
    month={Jul},
    day={06},
    volume={7},
    number={7},
    pages={e40530},
    issn={1932-6203},
    doi={10.1371/journal.pone.0040530},
    language={en}
}

@article{McManamay2022_HAI,
    title={Mapping hydrologic alteration and ecological consequences in stream reaches of the conterminous {United States}},
    author={McManamay, Ryan A and George, Rob and Morrison, Ryan R and Ruddell, Benjamin L},
    journal={Scientific Data},
    volume={9},
    number={1},
    pages={450},
    year={2022},
    publisher={Nature Publishing Group UK London}
}

@article{Morris2019_coverage,
  title={Using simulation studies to evaluate statistical methods},
  author={Morris, Tim P and White, Ian R and Crowther, Michael J},
  journal={Statistics in medicine},
  volume={38},
  number={11},
  pages={2074--2102},
  year={2019},
  publisher={Wiley Online Library}
}

@article{Morrison2023_FloodplainDegradation,
    title={Degradation of floodplain integrity within the contiguous {United States}},
    author={Morrison, Ryan R and Simonson, Kira and McManamay, Ryan A and Carver, Dan},
    journal={Communications Earth \& Environment},
    volume={4},
    number={1},
    pages={215},
    year={2023},
    publisher={Nature Publishing Group UK London}
}

@techreport{Moulton2002_revised,
  title={Revised protocols for sampling algal, invertebrate, and fish communities as part of the {National Water-Quality Assessment Program}},
  author={Moulton II, Stephen R and Kennen, Jonathan and Goldstein, Robert M and Hambrook, Julie A},
  year={2002},
  institution={Geological Survey (US)}
}

@techreport{Nagel2017_NSIHydroData,
    title={{National Stream Internet hydrography} datasets for spatial-stream-network ({SSN}) analysis},
    author={Nagel, D and Wollrab, S and Parkes-Payne, S and Peterson, E and Isaak, D and Ver Hoef, J},
    institution={Rocky Mountain Research Station, US Forest Service Data Archive, Fort Collins, CO},
    year={2017}
}

@article{Olea2011_bootstrap,
    title={Generalized bootstrap method for assessment of uncertainty in semivariogram inference},
    author={Olea, Ricardo A and Pardo-Ig{\'u}zquiza, Eulogio},
    journal={Mathematical Geosciences},
    volume={43},
    pages={203--228},
    year={2011},
    publisher={Springer}
}

@book{PageBurr1991_fieldguide,
  title={A field guide to freshwater fishes: {North America} north of {Mexico}},
  author={Page, Lawrence M and Burr, Brooks M},
  year={1991},
  publisher={Houghton Mifflin Harcourt}
}

@Article{Paukert2021,
    author={Paukert, Craig
    and Olden, Julian D.
    and Lynch, Abigail J.
    and Breshears, David D.
    and Christopher Chambers, R.
    and Chu, Cindy
    and Daly, Margaret
    and Dibble, Kimberly L.
    and Falke, Jeff
    and Isaak, Dan
    and Jacobson, Peter
    and Jensen, Olaf P.
    and Munroe, Daphne},
    title={Climate Change Effects on {North American} Fish and Fisheries to Inform Adaptation Strategies},
    journal={Fisheries},
    year={2021},
    month={Sep},
    day={01},
    volume={46},
    number={9},
    pages={449-464},
    doi={10.1002/fsh.10668},
    language={en}
}

@article{Peterson2006_StreamWaterChemistry,
    title={Patterns of spatial autocorrelation in stream water chemistry},
    author={Peterson, Erin E and Merton, Andrew A and Theobald, David M and Urquhart, N Scott},
    journal={Environmental Monitoring and Assessment},
    volume={121},
    pages={571--596},
    year={2006},
    publisher={Springer}
}

@article{Peterson2014_STARS,
    title={{STARS}: An {ArcGIS} toolset used to calculate the spatial information needed to fit spatial statistical models to stream network data},
    author={Peterson, Erin and Ver Hoef, Jay},
    journal={Journal of Statistical Software},
    volume={56},
    pages={1--17},
    year={2014}
}

@manual{Peterson2024_SSNbler,
    title = {{SSNbler}: Assemble {SSN} objects in {R}},
    author = {Erin Peterson and Michael Dumelle and Alan Pearse and Dan Teleki and Jay M. {Ver Hoef}},
    year = {2024},
    organization = {{R} package version 1.1.1}
}

@Article{Radinger2017,
    author={Radinger, Johannes
    and Essl, Franz
    and H{\"o}lker, Franz
    and Hork{\'y}, Pavel
    and Slav{\'i}k, Ond{\v{r}}ej
    and Wolter, Christian},
    title={The future distribution of river fish: The complex interplay of climate and land use changes, species dispersal and movement barriers},
    journal={Global Change Biology},
    year={2017},
    month={Nov},
    volume={23},
    number={11},
    pages={4970-4986},
    doi={10.1111/gcb.13760},
    language={en}
}

@book{RasmussenWilliams2006_GPs,
    title={Gaussian Processes for Machine Learning},
    author={Rasmussen, Carl Edward and Williams, Christopher KI},
    year={2006},
    publisher={MIT press},
    address={Cambridge, MA}
}

@article{Reid2019_emerging,
    title={Emerging threats and persistent conservation challenges for freshwater biodiversity},
    author={Reid, Andrea J and Carlson, Andrew K and Creed, Irena F and Eliason, Erika J and Gell, Peter A and Johnson, Pieter TJ and Kidd, Karen A and MacCormack, Tyson J and Olden, Julian D and Ormerod, Steve J and others},
    journal={Biological reviews},
    volume={94},
    number={3},
    pages={849--873},
    year={2019},
    publisher={Wiley Online Library}
}

@Article{Rogosch2019,
    author={Rogosch, Jane S.
    and Tonkin, Jonathan D.
    and Lytle, David A.
    and Merritt, David M.
    and Reynolds, Lindsay V.
    and Olden, Julian D.},
    title={Increasing drought favors nonnative fishes in a dryland river: evidence from a multispecies demographic model},
    journal={Ecosphere},
    year={2019},
    month={Apr},
    volume={10},
    number={4},
    pages={e02681},
    doi={10.1002/ecs2.2681},
    language={en}
}

@article{Saha2018_BRISC,
    title={{BRISC}: Bootstrap for rapid inference on spatial covariances},
    author={Saha, Arkajyoti and Datta, Abhirup},
    journal={Stat},
    volume={7},
    number={1},
    pages={e184},
    year={2018},
    publisher={Wiley Online Library}
}

@article{Saha2022_probitNNGP,
  title={Scalable predictions for spatial probit linear mixed models using nearest neighbor {Gaussian} processes},
  author={Saha, Arkajyoti and Datta, Abhirup and Banerjee, Sudipto},
  journal={Journal of data science: JDS},
  volume={20},
  number={4},
  pages={533},
  year={2022}
}

@article{Sainsbury2024_FRK,
    title={Modeling big, heterogeneous, {non-Gaussian} spatial and spatio-temporal data using {FRK}},
    author={Sainsbury-Dale, Matthew and Zammit-Mangion, Andrew and Cressie, Noel},
    journal={Journal of Statistical Software},
    volume={108},
    pages={1--39},
    year={2024}
}

@article{Schofield2018,
    author = {Schofield, Kate A. and Alexander, Laurie C. and Ridley, Caroline E. and Vanderhoof, Melanie K. and Fritz, Ken M. and Autrey, Bradley C. and DeMeester, Julie E. and Kepner, William G. and Lane, Charles R. and Leibowitz, Scott G. and Pollard, Amina I.},
    year = {2018},
    title = {Biota Connect Aquatic Habitats throughout Freshwater Ecosystem Mosaics},
    journal = {JAWRA Journal of the American Water Resources Association},
    volume = {54},
    number = {2},
    pages = {372-399},
    doi = {https://doi.org/10.1111/1752-1688.12634}
}

@book{Sedgewick2011_algorithms,
  author    = {Robert Sedgewick and Kevin Wayne},
  title     = {Algorithms},
  edition   = {4th},
  year      = {2011},
  publisher = {Addison-Wesley Professional},
  isbn      = {978-0-321-57351-3}
}

@article{SenguptaCressie2013_expfam,
  title={Hierarchical statistical modeling of big spatial datasets using the exponential family of distributions},
  author={Sengupta, Aritra and Cressie, Noel},
  journal={Spatial Statistics},
  volume={4},
  pages={14--44},
  year={2013},
  publisher={Elsevier}
}

@article{Stein2014_lowrank_limitations,
  title={Limitations on low rank approximations for covariance matrices of spatial data},
  author={Stein, Michael L},
  journal={Spatial Statistics},
  volume={8},
  pages={1--19},
  year={2014},
  publisher={Elsevier}
}

@article{Stein2002_screening,
  title={The screening effect in kriging},
  author={Stein, Michael L},
  journal={The Annals of Statistics},
  volume={30},
  number={1},
  pages={298--323},
  year={2002},
  publisher={Institute of Mathematical Statistics}
}

@article{Strecker2011_OldenLCRB,
    title={Defining conservation priorities for freshwater fishes according to taxonomic, functional, and phylogenetic diversity},
    author={Strecker, Angela L and Olden, Julian D and Whittier, Joanna B and Paukert, Craig P},
    journal={Ecological Applications},
    volume={21},
    number={8},
    pages={3002--3013},
    year={2011},
    publisher={Wiley Online Library}
}

@article{Tickner2020_GlobalFWBiodiversityLoss,
    author={Tickner, David
    and Opperman, Jeffrey J.
    and Abell, Robin
    and Acreman, Mike
    and Arthington, Angela H.
    and Bunn, Stuart E.
    and Cooke, Steven J.
    and Dalton, James
    and Darwall, Will
    and Edwards, Gavin
    and Harrison, Ian
    and Hughes, Kathy
    and Jones, Tim
    and Lecl{\`e}re, David
    and Lynch, Abigail J.
    and Leonard, Philip
    and McClain, Michael E.
    and Muruven, Dean
    and Olden, Julian D.
    and Ormerod, Steve J.
    and Robinson, James
    and Tharme, Rebecca E.
    and Thieme, Michele
    and Tockner, Klement
    and Wright, Mark
    and Young, Lucy},
    title={Bending the Curve of Global Freshwater Biodiversity Loss: An Emergency Recovery Plan},
    journal={BioScience},
    year={2020},
    month={Apr},
    day={01},
    volume={70},
    number={4},
    pages={330-342},
    doi={10.1093/biosci/biaa002},
    language={en}
}

@Article{Tonkin2021,
    author={Tonkin, Jonathan D.
    and Olden, Julian D.
    and Merritt, David M.
    and Reynolds, Lindsay V.
    and Rogosch, Jane S.
    and Lytle, David A.},
    title={Designing flow regimes to support entire river ecosystems},
    journal={Frontiers in Ecology and the Environment},
    year={2021},
    month={Aug},
    volume={19},
    number={6},
    pages={326-333},
    doi={10.1002/fee.2348},
    language={en}
}

@article{Urban2024_climatechange,
    author = {Mark C. Urban},
    title = {Climate change extinctions},
    journal = {Science},
    volume = {386},
    number = {6726},
    pages = {1123-1128},
    year = {2024},
    doi = {10.1126/science.adp4461}
}

@article{Vecchia1988_sparsity,
    title={Estimation and model identification for continuous spatial processes},
    author={Vecchia, Aldo V},
    journal={Journal of the Royal Statistical Society Series B: Statistical Methodology},
    volume={50},
    number={2},
    pages={297--312},
    year={1988},
    publisher={Oxford University Press}
}

@article{VerHoefBarry1996_MAcovar,
    title = {Constructing and fitting models for cokriging and multivariable spatial prediction},
    author = {Jay M. {Ver Hoef} and Ronald Paul Barry},
    journal = {Journal of Agricultural, Biological, and Environmental Statistics},
    volume = {1},
    number = {3},
    pages = {297-322},
    year = {1996},
    doi = {https://doi.org/10.2307/1400521}
}

@article{VerHoef2006_SpatialModelsThatUseFlow,
    title={Spatial statistical models that use flow and stream distance},
    author={Ver Hoef, Jay M and Peterson, Erin and Theobald, David},
    journal={Environmental and Ecological statistics},
    volume={13},
    pages={449--464},
    year={2006},
    publisher={Springer}
}

@article{VerHoef2010_MovingAverageApproach,
    title={A moving average approach for spatial statistical models of stream networks},
    author={Ver Hoef, Jay M and Peterson, Erin E},
    journal={Journal of the American Statistical Association},
    volume={105},
    number={489},
    pages={6--18},
    year={2010},
    publisher={Taylor \& Francis}
}

@article{VerHoef2014_SSN,
    title={{SSN}: An {R} package for spatial statistical modeling on stream networks},
    author={Ver Hoef, Jay and Peterson, Erin and Clifford, David and Shah, Rohan},
    journal={Journal of Statistical Software},
    volume={56},
    pages={1--45},
    year={2014}
}

@incollection{VerHoef2019_SpatialStreamModels,
    title={Spatial statistical models for stream networks},
    author={Ver Hoef, Jay M and Peterson, Erin E and Isaak, Daniel J},
    booktitle={Handbook of Environmental and Ecological Statistics},
    pages={421--444},
    year={2019},
    publisher={Chapman and Hall/CRC},
    editor={Gelfand, Alan E and Fuentes, Montserrat and Hoeting, Jennifer A and Smith, Richard Lyttleton}
}

@article{VerHoef2023_IndexingSPIN,
  title={Indexing and partitioning the spatial linear model for large data sets},
  author={Ver Hoef, Jay M and Dumelle, Michael and Higham, Matt and Peterson, Erin E and Isaak, Daniel J},
  journal={PLOS ONE},
  volume={18},
  number={11},
  pages={e0291906},
  year={2023},
  publisher={Public Library of Science San Francisco, CA USA}
}

@Article{Wagner2020,
    author={Wagner, Tyler
    and Hansen, Gretchen J.A.
    and Schliep, Erin M.
    and Bethke, Bethany J.
    and Honsey, Andrew E.
    and Jacobson, Peter C.
    and Kline, Benjamen C.
    and White, Shannon L.},
    title={Improved understanding and prediction of freshwater fish communities through the use of joint species distribution models},
    journal={Canadian Journal of Fisheries and Aquatic Sciences},
    year={2020},
    month={Sep},
    volume={77},
    number={9},
    pages={1540-1551},
    doi={10.1139/cjfas-2019-0348},
    language={en}
}

@article{Webster2008_Bayesian,
    title={Bayesian spatial modeling of data from unit-count surveys of fish in streams},
    author={Webster, Raymond A and Pollock, Kenneth H and Ghosh, Sujit K and Hankin, David G},
    journal={Transactions of the American Fisheries Society},
    volume={137},
    number={2},
    pages={438--453},
    year={2008},
    publisher={Taylor \& Francis}
}

@article{Wieczorek2018_covariateData,
    title={Select attributes for {NHDPlus} version 2.1 reach catchments and modified network routed upstream watersheds for the conterminous {United States}},
    author={Wieczorek, ME and Jackson, SE and Schwarz, GE},
    journal={US Geological Survey},
    volume={10},
    pages={F7765D7V},
    year={2018}
}

@article{Wyatt2003_Mapping,
    title={Mapping the abundance of riverine fish populations: integrating hierarchical {Bayesian} models with a geographic information system {(GIS)}},
    author={Wyatt, Robin J},
    journal={Canadian Journal of Fisheries and Aquatic Sciences},
    volume={60},
    number={8},
    pages={997--1006},
    year={2003},
    publisher={NRC Research Press Ottawa, Canada}
}

@book{Yaglom1987_MAcovar,
    title={Correlation Theory of Stationary and Related Random Functions},
    author={Yaglom, A.M.},
    isbn={9783540962687},
    lccn={86010167},
    year={1987},
    publisher={Springer}
}

@article{Ying1991_asymptotic,
  author  = {Ying, Zhiliang},
  title   = {Asymptotic properties of a maximum likelihood estimator with data from a {G}aussian process},
  journal = {Journal of Multivariate Analysis},
  volume  = {36},
  number  = {2},
  pages   = {280--296},
  year    = {1991}
}

@article{Zhang2004_asymptotic,
  author  = {Zhang, Hao},
  title   = {Inconsistent estimation and asymptotically equal interpolations in model-based geostatistics},
  journal = {Journal of the American Statistical Association},
  volume  = {99},
  number  = {465},
  pages   = {250--261},
  year    = {2004}
}

@article{ZilberKatzfuss2021_VecchiaLaplace,
  title={{Vecchia--Laplace} approximations of generalized {Gaussian} processes for big {non-Gaussian} spatial data},
  author={Zilber, Daniel and Katzfuss, Matthias},
  journal={Computational Statistics \& Data Analysis},
  volume={153},
  pages={107081},
  year={2021},
  publisher={Elsevier}
}

\end{document}